\documentclass[12pt,letterpaper]{article}
\pdfoutput=1
\usepackage[colorlinks=false,
linkcolor=red, 
citecolor=blue,
filecolor=red,
urlcolor=red,
linktoc=all, %%%
pdfstartview=FitV,
bookmarksopen=true]{hyperref}
\usepackage[left=2cm,top=1cm,right=3cm,nohead]{geometry}
\usepackage{xcolor,colortbl}
\usepackage[utf8]{inputenc} 
\usepackage{epsfig,latexsym,amsfonts,mathtools,amsthm,amssymb,amsbsy,multirow,slashed,color,
	mathrsfs,wasysym,textcomp,subfigure,wrapfig,comment,bbold,array,longtable,multirow,setspace}
\usepackage{tabularx}
\usepackage{graphicx}
\usepackage{setspace}
\usepackage{subfigure}
\usepackage[bf]{caption}
\usepackage{braket}
\usepackage{comment}
\usepackage{float}
\usepackage{tikz,pgfplots}
\usetikzlibrary{mindmap,trees,shadows}
\colorlet{color1}{gray!25}
\usetikzlibrary{positioning}
\usetikzlibrary{intersections} 

\newlength{\PicScale}
\usepackage[UKenglish]{babel}
\usepackage[toc,page]{appendix}
\usepackage{hhline}
\usepackage{geometry}
\usepackage{cite}
\usepackage{datetime}
\usepackage{bbm} 
\usepackage{bm} 
\hypersetup{
	pdftitle={},
	pdfauthor={},
	pdfsubject={}
} 

\definecolor{Gray}{gray}{0.94}
\newcommand{\mra}{{\mathrm{A}}}
\newcommand{\mrd}{{\mathrm{D}}}
\newcommand{\mre}{{\mathrm{E}}}
\newcommand{\mrc}{{\mathrm{C}}}
\newcommand{\mrb}{{\mathrm{B}}}
\newcommand{\mrf}{{\mathrm{F}}}
\newcommand{\mrg}{{\mathrm{G}}}

\newcommand{\rii}{{\mathrm{II}}}

\newcommand{\uo}{{\mathrm{U}(1)}}

\newcommand{\sug}{{\mathrm{SU}}}
\newcommand{\spg}{{\mathrm{Sp}}}
\newcommand{\sog}{{\mathrm{SO}}}
\newcommand{\sping}{{\mathrm{Spin}}}

\newcolumntype{M}[1]{>{\centering\arraybackslash}m{#1}}
\newcolumntype{N}{@{}m{0pt}@{}}

\numberwithin{equation}{section}
\renewcommand{\baselinestretch}{1.3}

\makeatletter
\def\@cline#1-#2\@nil{%%%
	\omit
	\@multicnt#1%%%
	\advance\@multispan\m@ne
	\ifnum\@multicnt=\@ne\@firstofone{&\omit}\fi
	\@multicnt#2%%%
	\advance\@multicnt-#1%%%
	\advance\@multispan\@ne
	\leaders\hrule\@height\arrayrulewidth\hfill
	\cr
	\noalign{\nobreak\vskip-\arrayrulewidth}}
\makeatother
\newtheorem{prop}{Proposition}

\graphicspath{{./Images/}}

\begin{document}
\pagestyle{empty}
\begin{center}        % Main title
  {\bf\LARGE Symmetry Enhancements in 7d \\ Heterotic Strings  \\ [3mm]}
%\vskip 0.1cm

\large{ Bernardo Fraiman$^{*\,\dagger}$ and H\'ector Parra De Freitas$^{*}$
 \\[2mm]}
{\small $*$  Institut de Physique Th\'eorique, Universit\'e Paris Saclay, CEA, CNRS\\ [-2mm]}
{\small\it  Orme des Merisiers, 91191 Gif-sur-Yvette CEDEX, France.\\[0.2cm] } 
{\small  $\dagger$ Instituto de Astronom\'ia y F\'isica del Espacio (IAFE-CONICET-UBA)\\ [-2mm]}
{\small \hspace{1em}  Departamento de F\'isica, FCEyN, Universidad de Buenos Aires (UBA) \\ [-2mm]}
{\small\it Ciudad Universitaria, Pabell\'on 1, 1428 Buenos Aires, Argentina\\ }

{\small \verb"bfraiman@iafe.uba.ar, hector.parradefreitas@ipht.fr"\\[-3mm]}
\vspace{0.3in}

\small{\bf Abstract} \\[3mm]\end{center}
We use a moduli space exploration algorithm to produce a complete list of maximally enhanced gauge groups that are realized in the heterotic string in 7d, encompassing the usual Narain component, and five other components with rank reduction realized via nontrivial holonomy triples. Using lattice embedding techniques we find an explicit match with the mechanism of singularity freezing in M-theory on K3. The complete global data for each gauge group is explicitly given.

\newpage

%\tableofcontents

%%%%%%%%%%%                 %%%%%%%%%%%%%%%%%%%
%%%%%%%%%%%  DOCUMENT BODY  %%%%%%%%%%%%%%%%%%%
%%%%%%%%%%%                 %%%%%%%%%%%%%%%%%%%

\setcounter{page}{1}
\pagestyle{plain}
\renewcommand{\thefootnote}{\arabic{footnote}}
\setcounter{footnote}{0}
%----------------------------------------------------------------------%
%  Paper begins
%----------------------------------------------------------------------%

%\tableofcontents
\tableofcontents	
\newpage
\section{Introduction}

The classification of gauge symmetries that can arise in the string landscape is an important problem that is comparatively easy to address for vacua with high supersymmetry. A particularly interesting set of vacua of this kind is provided by heterotic strings compactified on $T^d$, realizing models with 16 supercharges and gauge groups of rank $r = 16+d$. Although the basic mechanism governing the allowed gauge groups was determined a long time ago by Narain \cite{Narain:1985jj}, the full classification of all possibilities, together with their respective moduli, had not been made except for the case $d = 1$ \cite{Cachazo:2000ey} in the subsequent years. For $d = 2$ the list of possible gauge groups was implicitly known due to the duality between the heterotic string on $T^2$ and $F$-theory on elliptically fibered K3 surfaces, whose gauge groups were classified in \cite{SZ,Shimada2000OnEK}. 

Carrying out an exhaustive classification of allowed gauge groups in $d = 2$ was the main goal of \cite{Font:2020rsk}, where the results were found to match exactly with those of \cite{Shimada2000OnEK} but also the moduli for a representative model with each gauge group were given. It turned out that the most effective strategy for doing the classification consists on moving from singular points of maximal enhancement in moduli space to others via manipulations of the root lattices in a controlled way, in a process which is best described as an exploration algorithm. 

This algorithm is at heart a tool for finding embeddings of lattices into other lattices, which is precisely the context in which the mechanism of gauge symmetry enhancement is best formulated for the heterotic string on $T^d$. However it turns out that other components in the moduli space of theories with 16 supercharges can be treated in this way. By means of the asymmetric orbifold construction of \cite{Narain:1986qm}, theories with 16 supercharges but with gauge symmetry with reduced rank can be obtained. In particular, one finds the so called CHL string component \cite{Chaudhuri:1995fk,Chaudhuri:1995bf} in 9d, for which the momentum lattice was constructed by Mikhailov in \cite{Mikhailov:1998si}. In 8d the story is the same, but in 7d one finds four extra components (six including the Narain and CHL component). These were constructed together with their momentum lattices in \cite{deBoer:2001wca}. Classifying the possible gauge groups in the CHL string by means of the exploration algorithm was the main goal of \cite{Font:2021uyw}, where the respective topologies of the groups were obtained using results of \cite{Cvetic:2021sjm}.

In this paper we extend this work to the six aforementioned components in 7d. To this end it is necessary to state precisely how the enhanced symmetry groups can be obtained from the momentum lattices, which we do by a natural generalization of the case for the CHL string \cite{Mikhailov:1998si,Font:2021uyw,Cvetic:2021sjm}. We see that the lattice alone is not sufficient to determine the allowed gauge groups, but rather one must impose a constraint in the embeddings characterized by an integer, which comes from the string theory but is ad hoc from the point of view of the lattice (see Proposition \ref{propTrip}). Implementing this constraint in our algorithm we obtain a list of maximally enhanced gauge algebras for each component. We find 1035, 407, 50, 16, 3, and 3 such algebras for the $\mathbb{Z}_m$-triples, respectively with $m = 1, 2, 3, 4, 5, 6$. If we distinguish the enhancements by the global data of the gauge group, the corresponding numbers are 1232, 429, 52, 18, 3 and 3.

On the other hand it is well known that the heterotic string on $T^3$ is dual to M-theory on K3. Gauge groups with reduced rank are realized in the later when there are so-called partially frozen singularities on the K3 \cite{deBoer:2001wca,Atiyah:2001qf,Tachikawa:2015wka}. It is then natural to ask how this mechanism of partial freezing appears in the heterotic string. We study this problem by exploiting relations between the reduced rank momentum lattices and the Narain lattice and find a match with the known results in the M-theory side. General freezing rules involving the topology of the gauge groups are obtained, generalizing the results of \cite{Cvetic:2021sjm} for the 8d CHL string.

This paper is organized as follows. In Section \ref{s:review} we review the construction of rank reduced heterotic theories in nine to seven dimensions, emphasizing the role of outer automorphisms of the gauge lattice in the framework of asymmetric orbifolds. Then in Section \ref{s:lattices} we state the criteria for gauge groups being realized in the relevant theories in terms of lattice embeddings, and briefly review how the exploration algorithm works. The problem of singularity freezing is studied in Section \ref{s:frozen}. Finally, the main results obtained with the exploration algorithm are presented and discussed in Section \ref{s:results}. In Appendix \ref{app} we leave some comments regarding the role of some technicalities of lattice embeddings in the heterotic string, which may help the reader who is not used to thinking in these terms.

\section{Basic constructions with rank reduction}
\label{s:review}
In this section we review how rank reduced theories with 16 supercharges are constructed from the heterotic string in nine to seven dimensions. The idea is to get an intuitive understanding of these constructions through the manipulation of Dynkin Diagrams, illustrating the asymmetric orbifold construction with an outer automorphism. This complements the more general (and abstract) treatment in \cite{deBoer:2001wca}. We go through the CHL string, the $\sping(32)/\mathbb{Z}_2$ heterotic theory compactification without vector structure and the $\mathbb{Z}_m$-triples. %The later sections where we present our main results do not rely on this review; it may be safely skipped.  

\subsection{CHL string}
\label{chl}
The CHL string in 9d can be realized as the $\mre_8 \times \mre_8$ heterotic string compactified on an orbifold of a circle involving the outer automorphism $\theta$ which exchanges both $\mre_8$'s and a half-period shift $a$ along the circle \cite{Chaudhuri:1995bf}. The resulting target space has an holonomy $\theta$ along the compact direction which breaks the gauge group $\mre_8 \times \mre_8$ to its diagonal $\mre_8$. The shift $a$ obstructs the recovery of the broken $\mre_8$ in the twisted sector and so it ensures that the rank of the total gauge group is reduced.

Since $\theta$ is an outer automorphism of a gauge group, its implementation as an orbifold symmetry naturally leads to a picture of Dynkin Diagram folding. In the case of the CHL string, one ``folds one $\mre_8$ into the other", and finds that the gauge group of the resulting theory is $\mre_8$ (with an extra $\uo$ for arbitrary radius). Turning on a Wilson line does not change this picture since it must break both $\mre_8$'s in the same way, and one then just folds one of the broken groups into the other. 

Even though the length of a root is not by itself a meaningful concept, it is helpful to think that the nodes that get superposed in folding a diagram correspond to shortened roots. The reason is that this maps naturally to an increase in the level of the associated gauge algebra by a factor equal to the order of the automorphism $\theta$. In this case, the $\mre_8 \times \mre_8$ at level 1 becomes an $\mre_8$ at level 2. On the other hand, connected diagrams containing invariant nodes correspond to algebras at level 1. In the 9d CHL string there are no states in the gauge sector invariant under the orbifold symmetry, and so there are no gauge groups at level 1. Compactifying on a circle to 8d, one gets an extra $\text{SU(2)}$ at the self-dual radius which is unaffected by the folding and finds that indeed there are level 1 gauge symmetries (namely symplectic algebras of rank $\leq 10$).

The main idea here is that using the symmetry $(a,\theta)$ one constructs a vacuum of the heterotic string with an holonomy that in particular projects out Cartan generator states. Such an holonomy can not be implemented in the theory by merely turning on Wilson lines, as outer automorphisms are not connected to the identity element in the gauge group. However, the set of holonomies that can be obtained by orbifolding the target manifold is larger and includes those of this type. Together with the diagram folding picture, this story generalizes to the other constructions reviewed below.

\subsection{Compactification without vector structure}

There is a theory dual to the 8d CHL string which is obtained from the $\sping(32)/\mathbb{Z}_2$ heterotic string by compactifying it on a $T^2$ without vector structure \cite{Witten:1997bs}. The basic idea is that the spectrum of the 10d theory does not contain vector representations of $\sping(32)$, and so one should consider topologies of the gauge bundle which do not admit such representations. An obstruction of this type is measured by a mod two cohomology class $\tilde w_2$, analogous to the second Stieffel-Whitney class $w_2$ which obstructs spin structure. 

This compactification is characterized by the fact that the two holonomies $g_1, g_2$ on the torus commute as elements of $\sping(32)/\mathbb{Z}_2$, but do \textit{not} commute when lifted to elements of the double cover $\sping(32)$. In other words, the commutator of these holonomies is lifted to a nontrivial element in $\sping(32)$ which is identified with the identity upon quotienting by one of the spinor classes in its center. The lifting $\sping(32)/\mathbb{Z}_2 \to \sping(32)$  is therefore obstructed and no vector representations are allowed.

Two such holonomies can not be put simultaneously on a maximal torus of the gauge group. Similarly to the CHL string, one of them has to be realized by orbifolding the theory. The difference in this case is that the 10d gauge group $\sping(32)/\mathbb{Z}_2$ does not have any outer automorphism. One can however turn on a Wilson line along one of the compact directions such that from the point of view of the remaining dimensions the gauge group is actually broken to one which does in fact have an outer automorphism. Concretely, we turn on a Wilson line $A = (\tfrac12^8,0^8)$ which breaks $\sping(32)/\mathbb{Z}_2 \to \sping(16)^2/\mathbb{Z}_2$. This can be represented diagrammatically as
\begin{equation}\label{diag1}
\begin{tikzpicture}[scale = 0.9]
\draw(0,0)--(7,0);
%\draw(0.5,0)--(0.5,0.5);
\draw(6.5,0)--(6.5,0.5);
\draw(6.5,0.5)--(6.5,1);
\draw[fill=white](0,0) circle (0.1);
\draw[fill=white](0.5,0) circle (0.1);
\draw[fill=white](1,0) circle (0.1);
\draw[fill=white](1.5,0) circle (0.1);
\draw[fill=white](2,0) circle (0.1);
\draw[fill=white](2.5,0) circle (0.1);
\draw[fill=white](3,0) circle (0.1);
\draw[fill=white](3.5,0) circle (0.1);
\draw[fill=white](4,0) circle (0.1);
\draw[fill=white](4.5,0) circle (0.1);
\draw[fill=white](5,0) circle (0.1);
\draw[fill=white](5.5,0) circle (0.1);
\draw[fill=white](6,0) circle (0.1);
\draw[fill=white](6.5,0) circle (0.1);
\draw[fill=white](7,0) circle (0.1);
%\draw[fill=white](0.5,0.5) circle (0.1);
\draw[fill=white](6.5,0.5) circle (0.1);
\draw[fill=black](6.5,1) circle (0.1);
%
%\draw[red,->,>=stealth](3.25,-0.5)--(3.75,0.5) node[right]{$A$};
%
\draw[red,->,>=stealth](7.5,0.5)--(8.5,0.5) node[above=0.3,left]{$A$};
\draw[blue,->,>=stealth](7.5,-1.25)--(8.5,-1.25) node[above=0.3,left]{$\theta$};
\begin{scope}[shift={(9,0)}]
\draw(0,0)--(3,0);
\draw(4,0)--(7,0);
\draw(0.5,0)--(0.5,0.5);
\draw(6.5,0)--(6.5,0.5);
\draw(0.5,0.5)--(3.5,1);
\draw(6.5,0.5)--(3.5,1);
\draw[fill=white](0,0) circle (0.1);
\draw[fill=white](0.5,0) circle (0.1);
\draw[fill=white](1,0) circle (0.1);
\draw[fill=white](1.5,0) circle (0.1);
\draw[fill=white](2,0) circle (0.1);
\draw[fill=white](2.5,0) circle (0.1);
\draw[fill=white](3,0) circle (0.1);
%\draw[fill=white](3.5,0) circle (0.1);
\draw[fill=white](4,0) circle (0.1);
\draw[fill=white](4.5,0) circle (0.1);
\draw[fill=white](5,0) circle (0.1);
\draw[fill=white](5.5,0) circle (0.1);
\draw[fill=white](6,0) circle (0.1);
\draw[fill=white](6.5,0) circle (0.1);
\draw[fill=white](7,0) circle (0.1);
\draw[fill=white](0.5,0.5) circle (0.1);
\draw[fill=white](6.5,0.5) circle (0.1);
\draw[fill=black](3.5,1) circle (0.1);
\end{scope}
\begin{scope}[shift={(11,-1.5)}]
\draw(0,0)--(3,0);
\draw(0.5,0)--(0.5,0.5);
\draw[fill=white](0,0) circle (0.1);
\draw[fill=white](0.5,0) circle (0.1);
\draw[fill=white](1,0) circle (0.1);
\draw[fill=white](1.5,0) circle (0.1);
\draw[fill=white](2,0) circle (0.1);
\draw[fill=white](2.5,0) circle (0.1);
\draw[fill=white](3,0) circle (0.1);
\draw[fill=white](0.5,0.5) circle (0.1);
\end{scope}
\end{tikzpicture}
\end{equation}
where the white nodes are simple roots and the black nodes represent the fundamental weight which generates the $\mathbb{Z}_2$ in each case. We see that the RHS corresponds to a group with outer automorphism $\theta$. Orbifolding the theory by this symmetry and a half period shift along the second compact direction  we obtain a theory with gauge group $\sping(16) \times \uo^2$ (for arbitrary values of the torus metric and B-field). We note that the fundamental weight gets projected out by the orbifold symmetry, %hence the resulting gauge group is simply connected
but the gauge group is $\sping(16)/\mathbb{Z}_2$ \cite{Font:2021uyw,Cvetic:2021sjm}. 

The commutator of the holonomies chosen is the exponential of
\begin{equation}\label{comcond}
	A - \theta(A) = (\tfrac12^8,0^8) - (0^8,\tfrac12^8) = (\tfrac12^8,-\tfrac12^8),
\end{equation}
which does not yield the identity in $\sping(32)$ but rather the element which gets identified with it in $\sping(32)/\mathbb{Z}_2$. This corresponds to the discussion above. More generally one can deform this Wilson line by adding vectors symmetric in the first and last eight components, i.e. those of the form $(\delta,\delta)$, as to respect condition \eqref{comcond}. One can also turn on another Wilson line $A'$ in the second compact direction such that $\theta(A') = A'$, since the product of two holonomies on the same direction should commute. Together with deformations of the metric and the B-field we reach other points in moduli space exhibiting different gauge symmetries (classified in \cite{Font:2021uyw}). This moduli space is equivalent to that of the 8d CHL string, where the equivalence is given by T-duality\cite{deBoer:2001wca}.

\subsection{Holonomy triples in 7d}
\label{ss:triples}
The basic idea behind the construction just described can be applied to the heterotic string on a circle and further compactifying two dimensions on a torus. This comes from the fact that there are various 9d gauge groups analogous to the 10d $\sping(32)/\mathbb{Z}_2$. It is enough to consider the following five:
\begin{equation}
\footnotesize
\frac{(\mre_7 \times \sug(2))^2}{\mathbb{Z}_2}, ~~~ \frac{(\mre_6 \times \sug(3))^2}{\mathbb{Z}_3}, ~~~ \frac{(\sping(10) \times \sug(4))^2}{\mathbb{Z}_4}, ~~~ \frac{\sug(5)^4}{\mathbb{Z}_5}, ~~~\frac{(\sug(2) \times \sug(3) \times \sug(6))^2}{\mathbb{Z}_6}\,.
\end{equation}
These correspond to breakings of $\mre_8 \times \mre_8$ by a Wilson line $A$, so that it is most natural to work in the framework of the $\mre_8 \times \mre_8$ string. Natural choices for these Wilson lines are, respectively, 
\begin{equation}\label{1sthol}
A = 
	\begin{cases}
	(0^6,-\tfrac12,\tfrac12)\times(\tfrac12,-\tfrac12,0^6)\, & \quad(\mathbb{Z}_2)\\
	(0^5,-\tfrac13^2,\tfrac23)\times(\tfrac23,-\tfrac23^2,0^5)\, & \quad(\mathbb{Z}_3)\\
	(0^4,-\tfrac14^3,\tfrac34)\times (-\tfrac34,\tfrac14^3,0^4)\, & \quad(\mathbb{Z}_4)\\
	(0^3,-\tfrac15^4,\tfrac45)\times(-\tfrac45,\tfrac15^4,0^3)\, & \quad(\mathbb{Z}_5)\\
	(0^2,-\tfrac16^5,\tfrac56)\times(-\tfrac56,\tfrac16^5,0^2)\, & \quad(\mathbb{Z}_6)
	\end{cases}\,.
\end{equation}
The $\mathbb{Z}_m$'s correspond not only to the fundamental group of each broken gauge group but also to the cyclic group generated by the outer automorphism $\theta$ to be implemented. The name `$\mathbb{Z}_m$-triple' refers to this group together with the three holonomies consisting of \eqref{1sthol} and the pair analogous to the one discussed in the previous section, which we now discuss.
\subsubsection{\texorpdfstring{$\mathbb{Z}_2$}{Z2}-triple}
\label{ss:z2trip}
Consider first the $\mathbb{Z}_2$-triple. From the point of view of the $T^2$ on which the 9d theory is compactified, the gauge group is $(\mre_7 \times \sug(2))^2/\mathbb{Z}_2$, which indeed has an order two outer automorphism, exchanging the $\mre_7 \times \sug(2)$ factors. However, using this symmetry to orbifold the theory just gives us the CHL string, as discussed in section \ref{chl}. Consider instead turning on a Wilson line $A'$ on one of the $T^2$ directions ($x^1$), of the form
\begin{equation}
A' = (0^5,-\tfrac12,\tfrac12,0)\times (0,-\tfrac12,\tfrac12,0^5)\,.
\end{equation}
It has the effect of further breaking the gauge group to $(\mre_6\times \uo^2)^2$. From the point of view of the other $T^2$ direction ($x^2$), the gauge group has then an order 2 outer automorphism corresponding to the symmetry of each $\mre_6$ diagram. To get a consistent theory (meaning that the partition function is modular invariant), however, we have to take into account how the orbifold symmetry acts on the 16 internal directions and not only the 12 corresponding to the $\mre_6$'s. Fortunately, it is not hard to find such a consistent automorphism. One just has to take the one corresponding to the symmetry of the affine diagram of the original gauge algebra $2\mre_7 + 2\mra_1$:
\begin{equation}
\begin{tikzpicture}[scale = 0.9]
\draw(0,0)--(2,0);
\draw(0.5,0)--(0.5,1.5);
\draw(2.9,0)--(2.9,0.5);
\draw(3.1,0)--(3.1,0.5);
\draw[fill=white](0,0) circle (0.1);
\draw[fill=white](0.5,0) circle (0.1);
\draw[fill=white](1,0) circle (0.1);
\draw[fill=white](1.5,0) circle (0.1);
\draw[fill=white](2,0) circle (0.1);
%\draw[fill=white](2.5,0) circle (0.1);
\draw[fill=white](3,0) circle (0.1);
\draw[fill=white](3,0.5) circle (0.1);
%\draw[fill=white](3.5,0) circle (0.1);
%\draw[fill=white](4.5,0) circle (0.1);
\draw[fill=white](0.5,0.5) circle (0.1);
\draw[fill=white](0.5,1) circle (0.1);
\draw[fill=white](0.5,1.5) circle (0.1);
\draw[blue,<->,>=stealth](1.5,0.5) arc (0:90:0.5);
\draw[blue,<->,>=stealth](2.7,0)--(2.7,0.5);

\begin{scope}[xscale = -1, xshift = -7cm]
\draw(0,0)--(2,0);
\draw(0.5,0)--(0.5,1.5);
\draw(2.9,0)--(2.9,0.5);
\draw(3.1,0)--(3.1,0.5);
\draw[fill=white](0,0) circle (0.1);
\draw[fill=white](0.5,0) circle (0.1);
\draw[fill=white](1,0) circle (0.1);
\draw[fill=white](1.5,0) circle (0.1);
\draw[fill=white](2,0) circle (0.1);
%\draw[fill=white](2.5,0) circle (0.1);
\draw[fill=white](3,0) circle (0.1);
\draw[fill=white](3,0.5) circle (0.1);
%\draw[fill=white](3.5,0) circle (0.1);
%\draw[fill=white](4.5,0) circle (0.1);
\draw[fill=white](0.5,0.5) circle (0.1);
\draw[fill=white](0.5,1) circle (0.1);
\draw[fill=white](0.5,1.5) circle (0.1);
\draw[blue,<->,>=stealth](1.5,0.5) arc (0:90:0.5);
\draw[blue,<->,>=stealth](2.7,0)--(2.7,0.5);

\end{scope}
\end{tikzpicture}
\end{equation}
It can then be shown that, together with an order 2 shift in $x^2$, one obtains a consistent theory with an holonomy that breaks 8 Cartan generators, and the gauge group is $\mrf_4 \times \mrf_4$ at level 1 times $\uo^3$, for arbitrary metric and B-field. The former is due to the automorphism having an associated projector $P_\theta = 1 + \theta$ of rank 8. The later comes from the fact that each $\mre_6$ folds into an $\mrf_4$, where two nodes are left invariant (cf. discussion in section \ref{chl}). As in the previous construction, we can represent this breaking diagrammatically:
\begin{equation}
\begin{tikzpicture}[scale = 0.9]
\draw(0,0)--(2,0);
\draw(5,0)--(7,0);
\draw(0.5,0)--(0.5,1);
\draw(6.5,0)--(6.5,0.5);
\draw(6.5,0.5)--(6.5,1);
\draw(2,0)--(3.5,1);
\draw(3,0)--(3.5,1);
\draw(4,0)--(3.5,1);\\
\draw(5,0)--(3.5,1);
\draw[fill=white](0,0) circle (0.1);
\draw[fill=white](0.5,0) circle (0.1);
\draw[fill=white](1,0) circle (0.1);
\draw[fill=white](1.5,0) circle (0.1);
\draw[fill=white](2,0) circle (0.1);
%\draw[fill=white](2.5,0) circle (0.1);
\draw[fill=white](3,0) circle (0.1);
%\draw[fill=white](3.5,0) circle (0.1);
\draw[fill=white](4,0) circle (0.1);
%\draw[fill=white](4.5,0) circle (0.1);
\draw[fill=white](5,0) circle (0.1);
\draw[fill=white](5.5,0) circle (0.1);
\draw[fill=white](6,0) circle (0.1);
\draw[fill=white](6.5,0) circle (0.1);
\draw[fill=white](7,0) circle (0.1);
\draw[fill=white](0.5,0.5) circle (0.1);
\draw[fill=white](6.5,0.5) circle (0.1);
\draw[fill=white](6.5,1) circle (0.1);
\draw[fill=white](0.5,1) circle (0.1);
\draw[fill=black](3.5,1) circle (0.1);
%
%\draw[red,->,>=stealth](3.25,-0.5)--(3.75,0.5) node[right]{$A$};
%
\draw[red,->,>=stealth](7.5,0.5)--(8.5,0.5) node[above=0.3,left]{$A'$};
\draw[blue,->,>=stealth](7.5,-1.25)--(8.5,-1.25) node[above=0.3,left]{$\theta$};
\begin{scope}[shift={(9,0)}]
\draw(0,0)--(1.5,0);
\draw(5.5,0)--(7,0);
\draw(0.5,0)--(0.5,1);
\draw(6.5,0)--(6.5,0.5);
\draw(6.5,0.5)--(6.5,1);
\draw[fill=white](0,0) circle (0.1);
\draw[fill=white](0.5,0) circle (0.1);
\draw[fill=white](1,0) circle (0.1);
\draw[fill=white](1.5,0) circle (0.1);
\draw[fill=white](5.5,0) circle (0.1);
\draw[fill=white](6,0) circle (0.1);
\draw[fill=white](6.5,0) circle (0.1);
\draw[fill=white](7,0) circle (0.1);
\draw[fill=white](0.5,0.5) circle (0.1);
\draw[fill=white](6.5,0.5) circle (0.1);
\draw[fill=white](6.5,1) circle (0.1);
\draw[fill=white](0.5,1) circle (0.1);
\end{scope}
\begin{scope}[shift={(9,-1.5)}]
\draw(0,0)--(0.5,0);
\draw(0.5,0.05)--(1,0.05);
\draw(0.5,-0.05)--(1,-0.05);
\draw(1,0)--(1.5,0);
\draw(5.5,0)--(6,0);
\draw(6,0.05)--(6.5,0.05);
\draw(6,-0.05)--(6.5,-0.05);
\draw(6.5,0)--(7,0);
\draw[fill=white](0,0) circle (0.1);
\draw[fill=white](0.5,0) circle (0.1);
\draw[fill=white](1,0) circle (0.1);
\draw[fill=white](1.5,0) circle (0.1);
\draw[fill=white](5.5,0) circle (0.1);
\draw[fill=white](6,0) circle (0.1);
\draw[fill=white](6.5,0) circle (0.1);
\draw[fill=white](7,0) circle (0.1);
\end{scope}
\end{tikzpicture}
\end{equation}

Let us now consider the commutator of the holonomies along the $T^2$. We find that
\begin{equation}
	\theta(A') - A' = (0^5,1,-1,0)\times(0,1,-1,0^5)\,,
\end{equation}
which is just the fundamental weight represented as a black node in the above diagram. Its exponential is a nontrivial element of $(\mre_7 \times \sug(2)^2)$ which gets identified with the identity in the quotient $(\mre_7 \times \sug(2)^2)/\mathbb{Z}_2$, mirroring the situation in the compactification without vector structure as expected. One may also deform the Wilson lines along all directions by adding vectors invariant under $\theta$. This restriction reduces the degrees of freedom of the theory with respect to the Narain moduli space in the appropriate way. 

Finally we note that here we have obtained a particular gauge group, $\mrf_4 \times \mrf_4 \times \uo^3$, out of the many possibilities that exist in the moduli space of the theory. The general construction carried out in \cite{deBoer:2001wca} leads to a momentum lattice analogous to the Narain lattice, with which we may systematically explore this moduli space (as we discuss in next section). In this case, the momentum lattice is just the Mikhailov lattice in 7d and the theory is equivalent to the 7d CHL string. We emphasize that the $\mathbb{Z}_2$-triple does not involve the exchange of the $\mre_8$'s (or subgroups thereof), and so strictly speaking it does not correspond to the CHL string. Indeed, one can construct the CHL string but not the $\mathbb{Z}_2$-triple in 9d. When they exist, they are equivalent by T-duality.

\subsubsection{\texorpdfstring{$\mathbb{Z}_3$}{Z3}-triple}
Starting in the $\mathbb{Z}_3$-triple we find genuinely new rank-reduced moduli space components with respect to the 8d case. Here the gauge group from the point of view of the $T^2$ is $(\mre_6 \times \sug(3))^2/\mathbb{Z}_3$. We turn on a Wilson line along $x^1$ of the form
\begin{equation}
	A' = (0^4,-\tfrac13,\tfrac23,\tfrac13,0)\times (0,-\tfrac13,-\tfrac23,\tfrac13,0^4).
\end{equation}
This breaks the gauge group to $(\sog(8)\times \uo^4)^2$. To get the rank 3 automorphism we again consider the symmetry of the affine diagram of the original group:
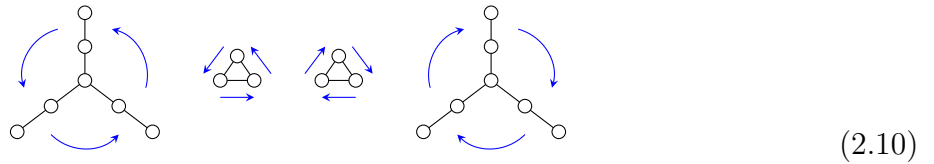
\begin{equation}
\begin{tikzpicture}[scale = 0.9]
\draw(-0.5,-0.76)--(0.5,0);
\draw(1.5,-0.76)--(0.5,0);
\draw(2.5,0)--(3,0);
\draw(0.5,0)--(0.5,1);
\draw(3,0)--(2.75,0.36);
\draw(2.5,0)--(2.75,0.36);
\begin{scope}[shift={(0,-0.05)}]
\draw[blue,->,>=stealth](0,-0.75)to[out=-45,in=225](1,-0.75);
\draw[blue,->,>=stealth,rotate around={120:(0.5,0)}](0,-0.75)to[out=-45,in=225](1,-0.75);
\draw[blue,->,>=stealth,rotate around={240:(0.5,0)}](0,-0.75)to[out=-45,in=225](1,-0.75);
\end{scope}
\draw[fill=white](0,-0.38) circle (0.1);
\draw[fill=white](0.5,0) circle (0.1);
\draw[fill=white](1,-0.38) circle (0.1);
\draw[fill=white](1.5,-0.76) circle (0.1);
\draw[fill=white](-0.5,-0.76) circle (0.1);
\draw[fill=white](2.5,0) circle (0.1);
\draw[fill=white](3,0) circle (0.1);
\draw[fill=white](2.75,0.36) circle (0.1);
%\draw[fill=white](3.5,0) circle (0.1);
%\draw[fill=white](4.5,0) circle (0.1);
\draw[fill=white](0.5,0.5) circle (0.1);
\draw[fill=white](0.5,1) circle (0.1);
%\draw[fill=white](0.5,1.5) circle (0.1);
%\draw[blue,->,>=stealth](1.25,0.25) arc (0:90:0.5);
%\draw[blue,->,>=stealth](0.25,0.75) arc (90:180:0.5);
%\draw[blue,->,>=stealth](0,-0.25) arc (180:360:0.5);
%\draw[blue,->,>=stealth](4,1.25) arc (0:315:0.5) node[left = 0.3, above]{$\frac{2\pi}{3}$°};

\draw[blue,->,>=stealth](2.55,0.5)--(2.25,0.1);
\draw[blue,->,>=stealth](2.5,-0.25)--(3,-0.25);
\draw[blue,->,>=stealth](3.25,0.1)--(2.95,0.5);

\begin{scope}[xscale = -1, xshift = -7cm]
\draw(-0.5,-0.76)--(0.5,0);
\draw(1.5,-0.76)--(0.5,0);
\draw(2.5,0)--(3,0);
\draw(0.5,0)--(0.5,1);
\draw(3,0)--(2.75,0.36);
\draw(2.5,0)--(2.75,0.36);
\begin{scope}[shift={(0,-0.05)}]
\draw[blue,->,>=stealth](0,-0.75)to[out=-45,in=225](1,-0.75);
\draw[blue,->,>=stealth,rotate around={120:(0.5,0)}](0,-0.75)to[out=-45,in=225](1,-0.75);
\draw[blue,->,>=stealth,rotate around={240:(0.5,0)}](0,-0.75)to[out=-45,in=225](1,-0.75);
\end{scope}
\draw[fill=white](0,-0.38) circle (0.1);
\draw[fill=white](0.5,0) circle (0.1);
\draw[fill=white](1,-0.38) circle (0.1);
\draw[fill=white](1.5,-0.76) circle (0.1);
\draw[fill=white](-0.5,-0.76) circle (0.1);
\draw[fill=white](2.5,0) circle (0.1);
\draw[fill=white](3,0) circle (0.1);
\draw[fill=white](2.75,0.36) circle (0.1);
%\draw[fill=white](3.5,0) circle (0.1);
%\draw[fill=white](4.5,0) circle (0.1);
\draw[fill=white](0.5,0.5) circle (0.1);
\draw[fill=white](0.5,1) circle (0.1);
%\draw[fill=white](0.5,1.5) circle (0.1);
%\draw[blue,->,>=stealth](1.25,0.25) arc (0:90:0.5);
%\draw[blue,->,>=stealth](0.25,0.75) arc (90:180:0.5);
%\draw[blue,->,>=stealth](0,-0.25) arc (180:360:0.5);
\draw[blue,->,>=stealth](2.55,0.5)--(2.25,0.1);
\draw[blue,->,>=stealth](2.5,-0.25)--(3,-0.25);
\draw[blue,->,>=stealth](3.25,0.1)--(2.95,0.5);
\end{scope}
\end{tikzpicture}
\end{equation}
This descends to the triality of each $\sog(8)$ and folds them into $\mrg_2 \times \mrg_2$ at level 1. The projector $P_\theta = 1 + \theta + \theta^2$ is of rank 4, eliminating 12 Cartan generators, and so the resulting gauge group is $\mrg_2 \times \mrg_2 \times \uo^3$ for arbitrary metric and B-field. Again, the orbifold includes an order 3 shift in $x^2$. The corresponding breaking diagram is 
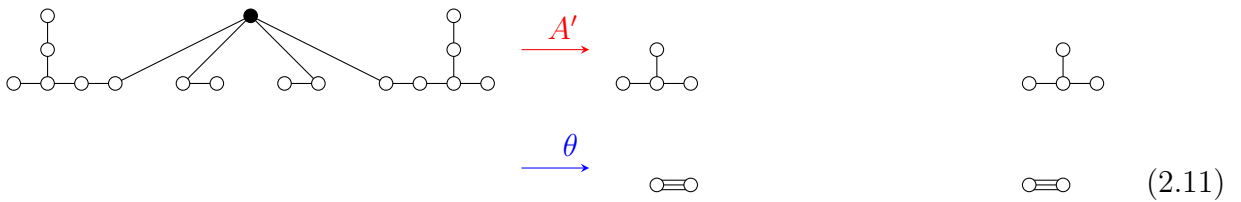
\begin{equation}
\begin{tikzpicture}[scale = 0.9]
\draw(0,0)--(1.5,0);
\draw(2.5,0)--(3,0);
\draw(4,0)--(4.5,0);
\draw(5.5,0)--(7,0);
\draw(0.5,0)--(0.5,1);
\draw(6.5,0)--(6.5,0.5);
\draw(6.5,0.5)--(6.5,1);
\draw(1.5,0)--(3.5,1);
\draw(2.5,0)--(3.5,1);
\draw(4.5,0)--(3.5,1);
\draw(5.5,0)--(3.5,1);
\draw[fill=white](0,0) circle (0.1);
\draw[fill=white](0.5,0) circle (0.1);
\draw[fill=white](1,0) circle (0.1);
\draw[fill=white](1.5,0) circle (0.1);
\draw[fill=white](2.5,0) circle (0.1);
\draw[fill=white](3,0) circle (0.1);
\draw[fill=white](4,0) circle (0.1);
\draw[fill=white](4.5,0) circle (0.1);
\draw[fill=white](5.5,0) circle (0.1);
\draw[fill=white](6,0) circle (0.1);
\draw[fill=white](6.5,0) circle (0.1);
\draw[fill=white](7,0) circle (0.1);
\draw[fill=white](0.5,0.5) circle (0.1);
\draw[fill=white](6.5,0.5) circle (0.1);
\draw[fill=white](6.5,1) circle (0.1);
\draw[fill=white](0.5,1) circle (0.1);
\draw[fill=black](3.5,1) circle (0.1);
\draw[red,->,>=stealth](7.5,0.5)--(8.5,0.5) node[above=0.3,left]{$A'$};
\draw[blue,->,>=stealth](7.5,-1.25)--(8.5,-1.25) node[above=0.3,left]{$\theta$};
\begin{scope}[shift={(9,0)}]
\draw(0,0)--(1,0);
\draw(6,0)--(7,0);
\draw(0.5,0)--(0.5,0.5);
\draw(6.5,0)--(6.5,0.5);
\draw[fill=white](0,0) circle (0.1);
\draw[fill=white](0.5,0) circle (0.1);
\draw[fill=white](1,0) circle (0.1);
\draw[fill=white](6,0) circle (0.1);
\draw[fill=white](6.5,0) circle (0.1);
\draw[fill=white](7,0) circle (0.1);
\draw[fill=white](0.5,0.5) circle (0.1);
\draw[fill=white](6.5,0.5) circle (0.1);
\end{scope}
\begin{scope}[shift={(9,-1.5)}]
\draw(0.5,0.07)--(1,0.07);
\draw(0.5,0)--(1,0);
\draw(0.5,-0.07)--(1,-0.07);
\draw(6,0.07)--(6.5,0.07);
\draw(6,0)--(6.5,0);
\draw(6,-0.07)--(6.5,-0.07);
\draw[fill=white](0.5,0) circle (0.1);
\draw[fill=white](1,0) circle (0.1);
\draw[fill=white](6,0) circle (0.1);
\draw[fill=white](6.5,0) circle (0.1);
\end{scope}
\end{tikzpicture}
\end{equation}

The commutator of $A'$ and $\theta$ is given by
\begin{equation}
	\theta(A')-A' =  (0^4,1,-1,0^2)\times(0^2,1,-1,0^4)\,,
\end{equation}
corresponding to the weight represented by the black node in the diagram above, and the story is the same as before for the $\mathbb{Z}_2$-triple. In this case one can deform the three Wilson lines with four degrees of freedom each, which is the rank of the projector $P_\theta$. Together with the nine degrees of freedom coming from the metric and B-field, the dimension of the moduli space is 21, and its local geometry is given by the coset
\begin{equation}
	\sog(7,3,\mathbb{R})\big/(\sog(7,\mathbb{R})\times \sog(3,\mathbb{R})).
\end{equation}  
In \cite{deBoer:2001wca} it was proposed that the global structure is given by the automorphism group of the momentum lattice of the theory, which was determined to be
\begin{equation}
	\Lambda_3 = \rii_{3,3}\oplus \mra_2 \oplus \mra_2,
\end{equation}
extending the results for the first two components of the moduli space where the Narain and the Mikhailov lattice respectively play this role. 

\subsubsection{\texorpdfstring{$\mathbb{Z}_4$}{Z4}-triple}\label{ss:z4}
For the $\mathbb{Z}_4$-triple we start with the 9d gauge group $(\sping(10)\times \sug(4))^2/\mathbb{Z}_4$ and turn on the Wilson line
\begin{equation}
	A' = \frac{1}{8}(1,-1,-1,-3,3,1,-1,5)\times(-5,1,-1,-3,3,1,1,-1),
\end{equation}
which breaks it to $\sug(3)^2 \times \uo^{12}$. The affine diagram of the original group has an order 4 symmetry:
\begin{equation}
\begin{tikzpicture}[scale = 0.9]
\draw(-0.5,-0.5)--(0,0);
\draw(0.5,0)--(1,-0.5);
\draw(-0.5,0.5)--(0,0);
\draw(0.5,0)--(1,0.5);
\draw(0,0)--(0.5,0);
\begin{scope}[shift={(0,-0.25)}]
\draw(2.5,0)--(3,0);
\draw(3,0)--(3,0.5);
\draw(3,0.5)--(2.5,0.5);
\draw(2.5,0)--(2.5,0.5);
\draw[fill=white](2.5,0.5) circle (0.1);
\draw[fill=white](2.5,0) circle (0.1);
\draw[fill=white](3,0) circle (0.1);
\draw[fill=white](3,0.5) circle (0.1);

\draw[blue,->,>=stealth](2.25,0.5)--(2.25,0);
\draw[blue,->,>=stealth](2.5,-0.25)--(3,-0.25);
\draw[blue,->,>=stealth](3.25,0)--(3.25,0.5);
\draw[blue,->,>=stealth](3,0.75)--(2.5,0.75);
\end{scope}
\draw[fill=white](0,0) circle (0.1);
\draw[fill=white](0.5,0) circle (0.1);
\draw[fill=white](1,-0.5) circle (0.1);
\draw[fill=white](-0.5,-0.5) circle (0.1);
\draw[fill=white](1,0.5) circle (0.1);
\draw[fill=white](-0.5,0.5) circle (0.1);
\draw[blue,->,>=stealth](0.75,0.75)to[out=135,in=45](-0.25,0.75);
\draw[blue,->,>=stealth](-0.5,0.25)to[out=270,in=180](0.75,-0.5);
\draw[blue,->,>=stealth](0.75,-0.75)to[out=-135,in=-45](-0.25,-0.75);
\draw[blue,->,>=stealth](-0.5,-0.25)to[out=-270,in=-180](0.75,0.5);
\draw[blue,<->,>=stealth](0,-0.25)--(0.5,-0.25);

\begin{scope}[xscale = -1, xshift = -7cm]
\draw(-0.5,-0.5)--(0,0);
\draw(0.5,0)--(1,-0.5);
\draw(-0.5,0.5)--(0,0);
\draw(0.5,0)--(1,0.5);
\draw(0,0)--(0.5,0);
\begin{scope}[shift={(0,-0.25)}]
\draw(2.5,0)--(3,0);
\draw(3,0)--(3,0.5);
\draw(3,0.5)--(2.5,0.5);
\draw(2.5,0)--(2.5,0.5);
\draw[fill=white](2.5,0.5) circle (0.1);
\draw[fill=white](2.5,0) circle (0.1);
\draw[fill=white](3,0) circle (0.1);
\draw[fill=white](3,0.5) circle (0.1);

\draw[blue,->,>=stealth](2.25,0.5)--(2.25,0);
\draw[blue,->,>=stealth](2.5,-0.25)--(3,-0.25);
\draw[blue,->,>=stealth](3.25,0)--(3.25,0.5);
\draw[blue,->,>=stealth](3,0.75)--(2.5,0.75);
\end{scope}
\draw[fill=white](0,0) circle (0.1);
\draw[fill=white](0.5,0) circle (0.1);
\draw[fill=white](1,-0.5) circle (0.1);
\draw[fill=white](-0.5,-0.5) circle (0.1);
\draw[fill=white](1,0.5) circle (0.1);
\draw[fill=white](-0.5,0.5) circle (0.1);
\draw[blue,->,>=stealth](0.75,0.75)to[out=135,in=45](-0.25,0.75);
\draw[blue,->,>=stealth](-0.5,0.25)to[out=270,in=180](0.75,-0.5);
\draw[blue,->,>=stealth](0.75,-0.75)to[out=-135,in=-45](-0.25,-0.75);
\draw[blue,->,>=stealth](-0.5,-0.25)to[out=-270,in=-180](0.75,0.5);
\draw[blue,<->,>=stealth](0,-0.25)--(0.5,-0.25);\underline{\draw(-0.5,-0.5)--(0,0);
\draw(0.5,0)--(1,-0.5);
\draw(-0.5,0.5)--(0,0);
\draw(0.5,0)--(1,0.5);
\draw(0,0)--(0.5,0);
\begin{scope}[shift={(0,-0.25)}]
\draw(2.5,0)--(3,0);
\draw(3,0)--(3,0.5);
\draw(3,0.5)--(2.5,0.5);
\draw(2.5,0)--(2.5,0.5);
\draw[fill=white](2.5,0.5) circle (0.1);
\draw[fill=white](2.5,0) circle (0.1);
\draw[fill=white](3,0) circle (0.1);
\draw[fill=white](3,0.5) circle (0.1);
\end{scope}
\draw[fill=white](0,0) circle (0.1);
\draw[fill=white](0.5,0) circle (0.1);
\draw[fill=white](1,-0.5) circle (0.1);
\draw[fill=white](-0.5,-0.5) circle (0.1);
\draw[fill=white](1,0.5) circle (0.1);
\draw[fill=white](-0.5,0.5) circle (0.1);
}
\end{scope}
\end{tikzpicture}
\end{equation}
The surviving $\sug(3)$'s under the action of $A'$ correspond to the innermost nodes of the affine $\sping(10)$'s, and they get identified under $\theta$ into $\sug(2)\times \sug(2)$ at level 1. The rank of the projector $P_\theta = 1 + \theta + \theta^2 + \theta^3$ is 2, and so 14 Cartan generators are eliminated. There is again an order 4 shift in $x^2$ in the orbifold symmetry, and we get the gauge group $\sug(2)\times \sug(2)\times \uo^3$ for generic metric and B-field. The breaking diagram is
\begin{equation}
\begin{tikzpicture}[scale = 0.9]
\draw(0,0)--(1,0);
\draw(2,0)--(3,0);
\draw(4,0)--(5,0);
\draw(6,0)--(7,0);
\draw(0.5,0)--(0.5,1);
\draw(6.5,0)--(6.5,0.5);
\draw(6.5,0.5)--(6.5,1);
\draw(1,0)--(3.5,1);
\draw(2,0)--(3.5,1);
\draw(5,0)--(3.5,1);
\draw(6,0)--(3.5,1);
\draw[fill=white](0,0) circle (0.1);
\draw[fill=white](0.5,0) circle (0.1);
\draw[fill=white](1,0) circle (0.1);
\draw[fill=white](2,0) circle (0.1);
\draw[fill=white](2.5,0) circle (0.1);
\draw[fill=white](3,0) circle (0.1);
\draw[fill=white](4,0) circle (0.1);
\draw[fill=white](4.5,0) circle (0.1);
\draw[fill=white](5,0) circle (0.1);
\draw[fill=white](6,0) circle (0.1);
\draw[fill=white](6.5,0) circle (0.1);
\draw[fill=white](7,0) circle (0.1);
\draw[fill=white](0.5,0.5) circle (0.1);
\draw[fill=white](6.5,0.5) circle (0.1);
\draw[fill=white](6.5,1) circle (0.1);
\draw[fill=white](0.5,1) circle (0.1);
\draw[fill=black](3.5,1) circle (0.1);

\draw[red,->,>=stealth](7.5,0.5)--(8.5,0.5) node[above=0.3,left]{$A'$};
\draw[blue,->,>=stealth](7.5,-1.25)--(8.5,-1.25) node[above=0.3,left]{$\theta$};
\begin{scope}[shift={(9,0)}]
\draw(0.5,0)--(0.5,0.5);
\draw(6.5,0)--(6.5,0.5);
\draw[fill=white](0.5,0) circle (0.1);
\draw[fill=white](0.5,0.5) circle (0.1);
\draw[fill=white](6.5,0) circle (0.1);
\draw[fill=white](6.5,0.5) circle (0.1);
\end{scope}
\begin{scope}[shift={(9,-1.5)}]
\draw[fill=white](1,0) circle (0.1);
\draw[fill=white](6,0) circle (0.1);
\end{scope}
\end{tikzpicture}
\end{equation}
We remark that the roots obtained after the folding have norm 8, this being the reason that the $\sug(2)$'s are at level 1. This can be understood by noting that the affine diagram for $\mrd_5$ gets folded into a pair of linked nodes with norms 2 and 8, respectively. Four nodes collapse into one corresponding to a root with norm smaller by a factor of 4, while two linked nodes fold into one with invariant length. Upon scaling, the shorter root that gets broken is of norm 2, while the remaining has norm 8.

We find that
\begin{equation}
	\theta(A')-A' = (0^3,1,0^3,-1)\times (1,0^3,-1,
	0^3)\,,
\end{equation} 
which is the weight in the LHS of the diagram above modulo a translation in the $A_3$ sublattices. The moduli space is of dimension 15, locally of the form
\begin{equation}
\sog(5,3,\mathbb{R})\big/(\sog(5,\mathbb{R})\times \sog(3,\mathbb{R})),
\end{equation}  
and the momentum lattice is
\begin{equation}
	\Lambda_4 = \rii_{3,3} \oplus \mra_1 \oplus \mra_1.
\end{equation}

\subsubsection{\texorpdfstring{$\mathbb{Z}_{5}$}{Z5} and \texorpdfstring{$\mathbb{Z}_6$}{Z6}-triples} 
For the $\mathbb{Z}_5$-triple we use Wilson line
\begin{equation}
	A' = \frac{1}{5}(0,-1,-2,3,2,1,0,-1)\times(1,0,-1,-2,-3,2,1,0)\,,
\end{equation}
which breaks $\sug(5)^4/\mathbb{Z}_5$ to $\uo^{16}$. The automorphism $\theta$ corresponds to the symmetry
\begin{equation}
\begin{tikzpicture}[scale = 0.9]
\draw(0,0)--(1,0);
\draw(1,0)--(1.31,0.95);
\draw(1.31,0.95)--(0.51,1.54);
\draw(0.51,1.54)--(-0.31,0.95);
\draw(-0.31,0.95)--(0,0);

\draw[fill=white](0,0) circle (0.1);
\draw[fill=white](1,0) circle (0.1);
\draw[fill=white](1.31,0.95) circle (0.1);
\draw[fill=white](0.51,1.54) circle (0.1);
\draw[fill=white](-0.31,0.95) circle (0.1);

\draw[blue,->,>=stealth](0,-0.25)--(1,-0.25)node[below=0.3, left]{$\times~2$};
\draw[blue,->,>=stealth](1.23,-0.08)--(1.54,0.87);
\draw[blue,->,>=stealth](1.46,1.15)--(0.66,1.74);
\draw[blue,->,>=stealth](0.34,1.74)--(-0.46,1.15);
\draw[blue,->,>=stealth](-0.54,0.87)--(-0.23,-0.08);
\begin{scope}[shift={(3,0)}]
\draw(0,0)--(1,0);
\draw(1,0)--(1.31,0.95);
\draw(1.31,0.95)--(0.51,1.54);
\draw(0.51,1.54)--(-0.31,0.95);
\draw(-0.31,0.95)--(0,0);

\draw[fill=white](0,0) circle (0.1);
\draw[fill=white](1,0) circle (0.1);
\draw[fill=white](1.31,0.95) circle (0.1);
\draw[fill=white](0.51,1.54) circle (0.1);
\draw[fill=white](-0.31,0.95) circle (0.1);

\draw[blue,->,>=stealth](0,-0.25)--(1,-0.25);
\draw[blue,->,>=stealth](1.23,-0.08)--(1.54,0.87);
\draw[blue,->,>=stealth](1.46,1.15)--(0.66,1.74);
\draw[blue,->,>=stealth](0.34,1.74)--(-0.46,1.15);
\draw[blue,->,>=stealth](-0.54,0.87)--(-0.23,-0.08);
\end{scope}
\begin{scope}[xscale = -1 , xshift = -10cm]
\draw(0,0)--(1,0);
\draw(1,0)--(1.31,0.95);
\draw(1.31,0.95)--(0.51,1.54);
\draw(0.51,1.54)--(-0.31,0.95);
\draw(-0.31,0.95)--(0,0);

\draw[fill=white](0,0) circle (0.1);
\draw[fill=white](1,0) circle (0.1);
\draw[fill=white](1.31,0.95) circle (0.1);
\draw[fill=white](0.51,1.54) circle (0.1);
\draw[fill=white](-0.31,0.95) circle (0.1);

\draw[blue,->,>=stealth](0,-0.25)--(1,-0.25)node[below=0.3, right]{$\times~2$};
\draw[blue,->,>=stealth](1.23,-0.08)--(1.54,0.87);
\draw[blue,->,>=stealth](1.46,1.15)--(0.66,1.74);
\draw[blue,->,>=stealth](0.34,1.74)--(-0.46,1.15);
\draw[blue,->,>=stealth](-0.54,0.87)--(-0.23,-0.08);
\begin{scope}[shift={(3,0)}]
\draw(0,0)--(1,0);
\draw(1,0)--(1.31,0.95);
\draw(1.31,0.95)--(0.51,1.54);
\draw(0.51,1.54)--(-0.31,0.95);
\draw(-0.31,0.95)--(0,0);

\draw[fill=white](0,0) circle (0.1);
\draw[fill=white](1,0) circle (0.1);
\draw[fill=white](1.31,0.95) circle (0.1);
\draw[fill=white](0.51,1.54) circle (0.1);
\draw[fill=white](-0.31,0.95) circle (0.1);

\draw[blue,->,>=stealth](0,-0.25)--(1,-0.25);
\draw[blue,->,>=stealth](1.23,-0.08)--(1.54,0.87);
\draw[blue,->,>=stealth](1.46,1.15)--(0.66,1.74);
\draw[blue,->,>=stealth](0.34,1.74)--(-0.46,1.15);
\draw[blue,->,>=stealth](-0.54,0.87)--(-0.23,-0.08);
\end{scope}
\end{scope}
\end{tikzpicture}
\end{equation}
and has projector $P_\theta = 0$. The rank of the gauge group is reduced by a factor of 16 and only the Cartans coming from the $T^3$ compactification are present. We have that
\begin{equation}
\theta(A')-A' = (0^2,1,-1,0^4)\times (0^4,1,-1,0^2)\,,
\end{equation} 
which is the weight associated to the $\mathbb{Z}_5$ quotient. The moduli space has dimension 9 and is locally of the form
\begin{equation}
\sog(3,3,\mathbb{R})\big/(\sog(3,\mathbb{R})\times \sog(3,\mathbb{R})),
\end{equation}  
and the momentum lattice is just $\rii_{3,3}$.

The story for the $\mathbb{Z}_6$-triple is basically the same, the only differences being that the Wilson line used is
\begin{equation}
	A' = \frac{1}{12}(1,-5,7,5,3,1,-1,-3)\times(3,1,-1,-3,-5,7,5,-1),
\end{equation}
the automorphism $\theta$ corresponds to the symmetry of the affine $(\sug(2)\times\sug(3)\times \sug(6))^2$ diagram,
\begin{equation}
\begin{tikzpicture}[scale = 0.9]
\begin{scope}[shift={(0,0)},scale=0.67]
\draw(0,0)--(1,0);
\draw(1,0)--(1.5,0.866);
\draw(1.5,0.866)--(1,1.71);
\draw(1,1.71)--(0,1.71);
\draw(0,1.71)--(-0.5,0.866);
\draw(-0.5,0.866)--(0,0);
\draw[fill=white](0,0) circle (0.15);
\draw[fill=white](1,0) circle (0.15);
\draw[fill=white](1.5,0.866) circle (0.15);
\draw[fill=white](1,1.71) circle (0.15);
\draw[fill=white](0,1.71) circle (0.15);
\draw[fill=white](-0.5,0.866) circle (0.15);
\draw[blue,->,>=stealth](0,0-0.375)--(1,0-0.375);
\draw[blue,->,>=stealth](1+0.375,0-0.1875)--(1.5+0.375,0.866-0.1875);
\draw[blue,->,>=stealth](1.5+0.375,0.866+0.1875)--(1+0.375,1.71+0.1875);
\draw[blue,->,>=stealth](1,1.71+0.375)--(0,1.71+0.375);
\draw[blue,->,>=stealth](0-0.375,1.71+0.1875)--(-0.5-0.375,0.866+0.1875);
\draw[blue,->,>=stealth](-0.5-0.375,0.866-0.1875)--(0-0.375,0-0.1875);
\end{scope}
\begin{scope}[shift={(2.5,0)}]
\draw(0,0)--(.5,0);
\draw(.5,0)--(.25,0.36);
\draw(0,0)--(.25,0.36);
\draw[blue,->,>=stealth](.05,0.5)--(-.25,0.1);
\draw[blue,->,>=stealth](0,-0.25)--(.5,-0.25);
\draw[blue,->,>=stealth](.75,0.1)--(.45,0.5);
\draw[fill=white](0,0) circle (0.1);
\draw[fill=white](.5,0) circle (0.1);
\draw[fill=white](.25,0.36) circle (0.1);
\end{scope}
\begin{scope}[shift={(4.5,0)}]
\draw[fill=white](0,0) circle (0.1);
\draw[fill=white](0,0.5) circle (0.1);
\draw[blue,<->,>=stealth](-0.3,0)--(-0.3,0.5);
\draw(-0.1,0)--(-0.1,0.5);
\draw(0.1,0)--(0.1,0.5);
\end{scope}
\begin{scope}[shift={(5.5,0)}]
\draw[fill=white](0,0) circle (0.1);
\draw[fill=white](0,0.5) circle (0.1);
\draw[blue,<->,>=stealth](--0.3,0)--(--0.3,0.5);
\draw(--0.1,0)--(--0.1,0.5);
\draw(-0.1,0)--(-0.1,0.5);
\end{scope}
\begin{scope}[shift={(7.5,0)}]
\draw(0,0)--(-.5,0);
\draw(-.5,0)--(-.25,0.36);
\draw(0,0)--(-.25,0.36);
\draw[blue,->,>=stealth](-.05,0.5)--(--.25,0.1);
\draw[blue,->,>=stealth](-0,-0.25)--(-.5,-0.25);
\draw[blue,->,>=stealth](-.75,0.1)--(-.45,0.5);
\draw[fill=white](0,0) circle (0.1);
\draw[fill=white](-.5,0) circle (0.1);
\draw[fill=white](-.25,0.36) circle (0.1);
\end{scope}
\begin{scope}[shift={(10,0)},scale=0.67]
\draw(-0,0)--(-1,0);
\draw(-1,0)--(-1.5,0.866);
\draw(-1.5,0.866)--(-1,1.71);
\draw(-1,1.71)--(-0,1.71);
\draw(-0,1.71)--(--0.5,0.866);
\draw(--0.5,0.866)--(-0,0);
\draw[fill=white](-0,0) circle (0.15);
\draw[fill=white](-1,0) circle (0.15);
\draw[fill=white](-1.5,0.866) circle (0.15);
\draw[fill=white](-1,1.71) circle (0.15);
\draw[fill=white](-0,1.71) circle (0.15);
\draw[fill=white](--0.5,0.866) circle (0.15);
\draw[blue,->,>=stealth](-0,0-0.375)--(-1,0-0.375);
\draw[blue,->,>=stealth](-1-0.375,0-0.1875)--(-1.5-0.375,0.866-0.1875);
\draw[blue,->,>=stealth](-1.5-0.375,0.866+0.1875)--(-1-0.375,1.71+0.1875);
\draw[blue,->,>=stealth](-1,1.71+0.375)--(-0,1.71+0.375);
\draw[blue,->,>=stealth](-0--0.375,1.71+0.1875)--(--0.5--0.375,0.866+0.1875);
\draw[blue,->,>=stealth](--0.5--0.375,0.866-0.1875)--(-0--0.375,0-0.1875);
\end{scope}
\end{tikzpicture}
\end{equation}
and
\begin{equation}
\theta(A')-A' = (0,1,-1,0^5)\times (0^5,1,-1,0^1)\,.
\end{equation} 
As in the previous case there are no Wilson line degrees of freedom, and the local and global data for the moduli space are the same. One should note however that the groups which are realized at level 5 in the $\mathbb{Z}_5$-triple are realized in this case at level 6. Indeed, this information is not contained implicitly in the momentum lattice. 

\section{7d Heterotic String and Momentum Lattices}
\label{s:lattices}
Here we explain the basic machinery of how gauge symmetry groups can be obtained from the momentum lattices corresponding to certain 7d heterotic string compactifications with 16 supercharges. These include the Narain lattice for $T^3$ compactifications, the Mikhailov lattice for the 7d CHL string, and the four extra momentum lattices for components with further rank reduction obtained in \cite{deBoer:2001wca}.

\subsection{The Narain construction}
It was shown in \cite{Narain:1985jj} that the perturbative spectrum of the heterotic string on $T^d$ can be put in correspondence with an even self-dual Lorentzian lattice $\rii_{16+d,d}$ of signature $(+^{16+d},-^{d})$. This lattice is spanned by vectors $(P,p_L;p_R)$, where $P$ is the left gauge lattice momentum and $p_{L,R}$ are the right and left internal space momenta.

The only massless states in the spectrum have $p_R = 0$, and those which realize the adjoint representation of the gauge algebra $\mathfrak g$ also have $P^2 + p_L^2 = 2$. They correspond therefore to a set of length $\sqrt{2}$ vectors in $\rii_{16+d,d}$ spanning a positive definite sublattice $L$, which is just the root lattice of $\mathfrak g$. The question of what gauge algebras can be realized in the theory is then equivalent to the question of what root lattices $L$ can be embedded in the Narain lattice. Note that this embedding has to be such that the intersection of the real span of $L$ with $\rii_{16+d,d}$ does not contain a larger root lattice $L'$, since this would leave out extra states that do form part of the massless spectrum. 
We can be more precise about the relation between gauge symmetries and lattice embeddings and in the way gain more information. As discussed in \cite{Font:2021uyw}, relaxing the condition $P^2 + p_L^2 = 2$ while keeping $p_R = 0$ defines an overlattice $M \supseteq L$ corresponding to the weight lattice of the global gauge group $G$. In this case, $M$ is such that the intersection of its real span with $\rii_{16+d,d}$ is $M$ itself, i.e. it is \textit{primitively} embedded in $\rii_{16+d,d}$ (see Appendix \ref{app:prim} for an extended discussion regarding these embeddings). The full statement regarding the possibility of some gauge group $G$ being realized in the heterotic string on $T^d$ is as follows: 
\\
\\
\fbox{\parbox{\textwidth}{
\begin{prop}
Let $G = \tilde G/H$ be some semisimple group of rank $r \leq 16+d$, where $\tilde G$ and $H$ are respectively the universal cover and the fundamental group. $G \times U(1)^{16+d-r}$ is realized in the heterotic string on $T^d$ as a gauge symmetry group if and only if its weight lattice $M$ admits a primitive embedding in the Narain lattice $\rii_{16+d,d}$ such that the vectors in $M$ of length $\sqrt2$ are roots.
\end{prop}\label{propNar}
}}
\\
\\
At the end of the day, the classification of the possible gauge groups that can be obtained in the heterotic string on $T^d$ turns out to be a (conceptually) simple problem of lattice embeddings.   

\subsubsection{Exploration algorithm}
\label{sss:alg}

There exist many useful theorems and techniques, mainly due to Nikulin \cite{MR525944,Taylor:2011wt}, for determining if some lattice $M$ admits a primitive embedding in another lattice $\Lambda$. However, even if they can yield insight into the structure of the theory, they do not by themselves give an efficient method for obtaining a thorough classification of the allowed gauge groups. We are therefore led to develop more constructive methods which can be easily turned into computer algorithms.  

Such an algorithm was presented in \cite{Font:2020rsk}, and it works roughly as follows (see \cite{Font:2021uyw} for a detailed explanation): 

\begin{enumerate}
\item Take a point in moduli space of $T^d$ compactifications with maximally enhanced gauge group, say $\sping(32 + 2d)$, such that the embedding of the root lattice $L \hookrightarrow \rii_{16+d,d}$ is explicitly known.

\item Break the gauge group by removing a simple root. This relaxes a constraint on the moduli such that the semisimple rank reduced part of the gauge group generically corresponds to a $d$-dimensional subvariety $V$ of the moduli space.

\item Enhance the group with a different simple root than the one previously removed. This generically selects a different point of maximal enhancement contained in $V$.
\end{enumerate}
This procedure is repeated for different choices of breakings-enhancements for a given starting point, and then repeated again starting from the newly found gauge groups, up until it does not yield new results. It is natural to assume that all points of maximal enhancement in moduli space can be reached in this way, and it is in fact true for the cases $d = 1, 2$ \cite{Font:2020rsk}.

Non-maximal enhancements can be obtained from the maximal ones by simply removing an arbitrary number of roots. Remarkably, for $d = 1,2$ there are respectively only two gauge groups which can not be obtained in this way, namely $\sping(16)^2/\mathbb{Z}_2$ for $d = 1$ and $\sping(8)^4/\mathbb{Z}_2^2$ for $d = 2$. In $d = 3$ there is also such  gauge gauge group,
\begin{equation}\label{noenh}
    G = \frac{\sug(2)^{16}}{\mathbb{Z}_2^5}\,,
\end{equation}
where the fundamental group is given by \begin{equation}\def\arraystretch{0.7}
  \begin{footnotesize}\begin{array}{@{}c@{}c@{}c@{}c@{}c@{}c@{}c@{}c@{}c@{}c@{}c@{}c@{}c@{}c@{}c@{}c@{}}
0&	0&	0&	0&	0&	0&	0&	0&	1&	1&	1&	1&	1&	1&	1&	1\\
0&	0&	0&	0&	1&	1&	1&	1&	0&	0&	0&	0&	1&	1&	1&	1\\
0&	0&	1&	1&	0&	0&	1&	1&	0&	0&	1&	1&	0&	0&	1&	1\\
0&	1&	0&	1&	0&	1&	0&	1&	0&	1&	0&	1&	0&	1&	0&	1\\
1&	0&	0&	1&	0&	1&	1&	0&	0&	1&	1&	0&	1&	0&	0&	1\\
	\end{array}\end{footnotesize}
\end{equation}
At the level of gauge algebras this was already noted in \cite{Polchinski:1995df}.

In this paper we are interested in the possible gauge groups that can be realized in the heterotic string on $T^3$. Using the exploration algorithm just described, we have collected a set of points of maximal enhancement characterized by their root lattices $L$, i.e. their gauge algebras $\mathfrak g$. For each point we compute the weight lattice $M$ and from it the generators of the fundamental group $H$, using the methods described in \cite{Font:2021uyw}. The results are presented in Section \ref{s:results}.

\subsection{The CHL string and Mikhailov lattice}

Now we wish to extend the discussion of the previous subsection to the CHL string on $T^d$, which can be realized as an asymmetric orbifold of the heterotic string on $T^d$. The analog of the Narain lattice for this theory was constructed by Mikhailov in \cite{Mikhailov:1998si} and can be written as
\begin{equation}
	\rii_{(d)} = \rii_{d-1,d-1}(2)\oplus \rii_{1,1} \oplus \mre_8\,,
\end{equation}
where the $(2)$ indicates that $\rii_{d-1,d-1}$ is scaled by a factor of $\sqrt{2}$. Depending on the dimension $d$, this lattice may be rewritten in different ways using lattice isomorphisms. For $d = 3$, we have
\begin{equation}
	\rii_{2,2}(2)\oplus \rii_{1,1} \oplus \mre_8 ~~\simeq~~ \rii_{3,3} \oplus \mrd_4 \oplus \mrd_4 ~~\simeq~~ \rii_{3,3} \oplus  \mrf_4 \oplus \mrf_4 \,.
\end{equation}
Here we have used the root lattice isomorphism $\mrd_4 \simeq \mrf_4$ (the corresponding root \textit{systems} are of course not isomorphic, see Appendix \ref{app:iso}) to reflect the fact that the `canonical' point in the theory has gauge algebra $2\mrf_4$ and not $2 \mrd_4$, as shown in Section \ref{ss:z2trip}. 

The relation between lattice embeddings and realizability of gauge groups in the CHL string is more complicated than for the usual heterotic string on tori. In the latter, the roots of the gauge algebra correspond to the length $\sqrt2$ vectors in some positive definite lattice $\Lambda$ primitively embedded into $\rii_{16+d,d}$. In the CHL string the mass formulas are such that it is also possible for some but not all vectors of length $2$ to give roots. In order for such a vector $v$ to correspond to a root, it must satisfy the condition that its inner product with all other vectors in the whole Mikhailov lattice is even \cite{Mikhailov:1998si}. In this case we say that $v$ is a level $2$ vector (not to be confused with the level of the Kac-Moody algebra for the gauge group). More generally, a vector $v$ in a lattice $\Lambda$ is said to be at embedding level $\ell$ if the product of $v$ with every vector in $\Lambda$ is divisible by $\ell$. 

On the other hand, the statement that the global structure of the gauge group is given by the primitively weight overlattice $M$ does not generalize to the case where the momentum lattice is not self-dual and the gauge algebras are not of ADE type. The problem of obtaining this global data was studied in detail in \cite{Cvetic:2021sjm}. It was shown in particular that the fundamental group $\pi_1(G)$ of the gauge group $G$ is given by the quotient of the cocharacter lattice $M^\vee$ and the coroot lattice $L^\vee$ where the later is embedded in the dual momentum lattice $\rii_{(d)}^*$ and the former is the corresponding overlattice which is primitively embedded in $\rii_{(d)}^*$. 

One strategy to obtain all the possible gauge groups in the theory is to apply the exploration algorithm described above to the dual lattice $\rii_{(d)}^*$ (which usually has to be rescaled to be made even) and compute the lattices $L$ and $M$ the same way as for the Narain lattice, but dualizing the algebra $\mathfrak g \to \mathfrak g^\vee$ at the end. It can be shown that the embedding level condition for vectors to be roots are the same as for the original lattice $\rii_{(d)}$. This corresponds to the method employed in \cite{Font:2021uyw} to obtain the list of gauge groups for the CHL string in 8d. 

Having dealt with this subtlety, a statement generalizing proposition \ref{propNar} for the usual heterotic string to the CHL string on $T^d$ can be made as follows:
\\
\\
\fbox{\parbox{\textwidth}{
		\begin{prop}
			Let $G = \tilde G/H$ be some semisimple group of rank $r \leq d+8$, where $\tilde G$ and $H$ are respectively the universal cover and the fundamental group. $G \times U(1)^{d+8-r}$ is realized in the CHL string on $T^d$ as a gauge symmetry group if and only if the weight lattice $M^\vee$ of the dual group $G^\vee$ admits a primitive embedding in the dual Mikhailov lattice $\rii_{(d)}^*(2)$ such that the vectors in $M^\vee$ of length $\sqrt{2 \ell}$ at embedding level $\ell = 1,2$ in $\rii_{(d)}^*(2)$ belong to $L^\vee$.
		\end{prop}\label{propCHL}
}}
\\
\\
We see that the embedding level $\ell$ plays an important role in the theory, allowing to treat the problem of finding the possible gauge groups without reference to the string theory itself, as in the case of the original heterotic string.

Finally let us recall that the simple factors in $G$ have associated Kac-Moody algebras at level $\mathfrak m = 1,2$ where $2/\mathfrak m$ is the squared length of the corresponding longest root. For $d = 2$ there are only ADE groups at level 2 and symplectic groups at level 1 (including $\spg(1) = \sug(2)$). This moduli space was exhaustively explored in \cite{Font:2021uyw} using an extension of the algorithm discussed in Section \ref{sss:alg}. For $d = 3$ there are more interesting possibilities including $\mrb_3$ and $\mrf_4$ at level 1. 

\subsection{Momentum lattices from Triples}

Let us now turn to the $\mathbb{Z}_m$-triples reviewed in Section \ref{ss:triples}. The respective momentum lattices are given in Table \ref{tab:lattices}, where we also show the rank reduction of the respective gauge groups. Here again we have chosen to write the lattices in terms of the canonical point groups using the lattice isomorphisms $\mrd_4 \simeq \mrf_4$ and $\mra_2 \simeq \mrg_2$. We also record the frozen singularity for each lattice $\Lambda_m$, which in this context corresponds to the orthogonal complement of the embedding $\Lambda_m \hookrightarrow \rii_{19,3}$. This point is discussed in more detail in the next section.

\begin{table}[htb]
	\begin{center}
		\begin{tabular}{|c|c|c|c|}\hline
			$m$ & $\Lambda_m$ & Frozen Singularity &$r_-$\\ \hline
			1&$\rii_{3,3}\oplus \mre_8 \oplus \mre_8$&$\emptyset$&0\\  \hline
			2&$\rii_{3,3} \oplus  \mrf_4 \oplus \mrf_4$&$\mrd_4\oplus \mrd_4$&8\\ \hline
			3&$\rii_{3,3}\oplus \mrg_2 \oplus \mrg_2$&$\mre_6 \oplus \mre_6$&12\\ \hline
			4&$\rii_{3,3}\oplus \mra_1 \oplus \mra_1$&$\mre_7 \oplus \mre_7$&14\\ \hline
			5&$\rii_{3,3}$&$\mre_8\oplus \mre_8$&16\\\hline
			6&$\rii_{3,3}$&$\mre_8 \oplus \mre_8$&16 \\ \hline
			\end{tabular}
		\caption{Momentum lattices $\Lambda_m$ for the moduli spaces of heterotic $\mathbb{Z}_m$-triples. The gauge group rank for $m = 1$ is 19, which is just the Narain component. The case $m = 2$ is dual to but not the same as the CHL component \cite{deBoer:2001wca}. The frozen singularities correspond to the orthogonal complements of $\Lambda_m \hookrightarrow \rii_{19,3}$.}
		\label{tab:lattices}
	\end{center} 
\end{table}

It is natural to ask whether we can extend propositions \ref{propNar} and \ref{propCHL} to these lattices. An obvious ansatz is the following:
\\
\\
\fbox{\parbox{\textwidth}{
		\begin{prop}
			Let $G = \tilde G/H$ be some semisimple group of rank $r \leq r_m$, where $\tilde G$ and $H$ are respectively the universal cover and the fundamental group, and $r_m = 19, 11, 7, 5, 3, 3$ respectively for $m = 1,...,6$. $G \times U(1)^{19-r_m}$ is realized in the $\mathbb{Z}_m$-triple as a gauge symmetry group if and only if the weight lattice $M^\vee$ of the dual group $G^\vee$ admits a primitive embedding in the dual momentum lattice $\Lambda_m^*(m)$ such that the vectors in $M^\vee$ of length $\sqrt{2\ell}$ at embedding level $\ell = 1,m$ in $\Lambda_m^*(m)$ belong to $L^\vee$. Simple factors are realized at level $\mathfrak m = 2m/\alpha_\text{long}^2$, where $\alpha_\text{long}$ is a long root in $L \hookrightarrow \Lambda_m$.
		\end{prop}\label{propTrip}
}}
\\
\\
The key ingredient is that the vectors of length $\sqrt{2m}$ at embedding level $m$ correspond to massless states and give e.g. long roots for non-ADE gauge groups. This can in fact be explicitly proved in the particular construction used in \cite{deBoer:2001wca} to obtain the momentum lattices. This roughly corresponds to the fact that in this construction there is a rescaling by a factor of $\sqrt{m}$ involved, such that the product of long roots, coming from invariant states in the parent theory of the orbifold, with all other vectors is scaled by a factor of $m$. We will however confirm this for the general case by showing in Section \ref{s:frozen} that assuming this ansatz one can reproduce the mechanism of singularity freezing in the dual M-theory on K3 from the heterotic side.

%We note however that there is a subtlety in the case $m = 4$ that makes some gauge groups not conform fo Proposition \ref{propTrip}. The lattice $\Lambda_4 = \rii_{3,3} \oplus \mra_1 \oplus \mra_1$ has dual lattice $\Lambda_4^*(4) = \rii_{3,3}(4)\oplus \mra_1 \oplus \mra_1$. The ``canonical" maximal enhancement in the theory has root lattice $L = 5 \mra_1$, so that the coroot lattice $L^\vee$ found in $\Lambda_4^*(4)$ should be $L^\vee = 5 \mra_1 (4)$. The canonical point in this lattice, however, has lattice $L^\vee = 3 \mra_1 (4) \oplus 2\mra_1$, which does not match with the correct coroot lattice. Indeed, the fundamentally correct approach is not to naively explore the dual lattices in the same way as the original ones. One should instead explicitly look for the coroot lattices of those root lattices obtained from the original lattice and take every other vector in its overlattice as part of the cocharacter lattice. We note however that both procedures give the exact same results except in the $m = 4$ case where these two $\mra_1$'s are involved. 

An extension of the exploration algorithm used for the CHL string to these lattices is straightforward and produces the results presented in Section \ref{ss:restrip}. In Section \ref{ss:proj} we will see that these can be reproduced by applying an appropriate projection map to the Narain component.

\section{Frozen singularities from the heterotic side}
\label{s:frozen}
It was already noted by Mikhailov in \cite{Mikhailov:1998si} that the momentum lattice for the CHL string is primitively embedded in the Narain lattice such that its orthogonal complement corresponds to the frozen singularity on the dual F/M-theories on K3 (for $d = 2,3$, respectively). This observation was extended in \cite{deBoer:2001wca} to the $\mathbb{Z}_m$-triples in 7d. Here we make use of it together with Proposition \ref{propTrip} to determine precisely how the ADE singularities are partially frozen (usually to give non-ADE algebras) and recover the known ``freezing rules" on the K3 side.  

\subsection{Freezing rules in 8d}

Let us first demonstrate the general method of obtaining the freezing rules in the $d = 2$ case, which map gauge groups in the Narain component to the CHL component of the moduli space. 

We start by considering an embedding of the Mikhailov lattice $\rii_{2,2} \oplus \mrd_8 \simeq \rii_{2,2} \oplus \mrc_8$ into the Narain lattice $\rii_{18,2}$. This is done in practice by taking the orthogonal complement of any primitively embedded $\mrd_8$ lattice in $\rii_{18,2}$, which is unique modulo automorphisms of the later. We then consider in turn an embedding of a $\mrc_n$ root lattice in the Mikhailov sublattice (cf. Proposition \ref{propCHL}), which will therefore be also embedded in the Narain lattice, 
\begin{equation}
	\mrc_n \hookrightarrow \rii_{2,2} \oplus \mrc_8 \hookrightarrow \rii_{18,2}, ~~~~~ n \leq 10 \,.
\end{equation}
This will correspond to an embedding $\mrc_n \oplus \mrd_8 \hookrightarrow \rii_{18,2}$ which will however neither be primitive, nor conform to the rules of Proposition \ref{propNar} due to the long roots. It does however define an $(n+8)$-plane in the ambient space of $\rii_{18,2}$ which in turn defines some primitively embedded weight lattice $M$. One may chose to focus only on the root sublattice $L \subseteq W$, which is enough to make the comparison with singularity freezing in F-theory. In this case we find that to the naive embedding of $\mrc_n \oplus \mrd_8$ inherited from the Mikhailov lattice and the frozen singularity there corresponds an actual embedding
\begin{equation}
	\mrd_{n+8} \hookrightarrow \rii_{18,2},
\end{equation} 
which may require extra weights (but not roots) to be made primitive. With this we recover the freezing rule for F-theory on K3 in the reverse. Indeed, applying these rules to all the possible gauge algebras in the Narain component gives those in the CHL component \cite{Font:2021uyw,Hamada:2021bbz}.

\subsection{Freezing rules in 7d}

In 7d there are more possibilities for freezing singularities, each one defining a different momentum lattice as shown in Table \ref{tab:lattices}. The process outlined above can be repeated in this case and we obtain the following patterns
\begin{equation}\label{frules}
	\begin{split}
	m = 2: ~~~&~~~\mrc_p + \mrc_q \to \mrd_{p+4} + \mrd_{q+4}\,,~~~ p,q \geq 0\,,\\
	       ~~~&~~~\mrc_p + \mrf_q \to \mrd_{p+4} + \mre_{q+4}\,,~~~ p \geq 0,~q = 2,3,4\,,\\
	       ~~~&~~~\mrf_p + \mrf_q \to \mre_{p+4} + \mre_{q+4}\,,~~~ p,q = 2,3,4\,,\\
	m = 3: ~~~&~~~\mrg_p + \mrg_q \to \mre_{6+p} + \mre_{6+q}\,,~~~ p,q = 0,1,2\,,\\
	m = 4: ~~~&~~~\mra_p + \mra_q \to \mre_{7+p} + \mre_{7+q}\,, ~~~ p,q = 0,1\,,\\
        %m = 4: ~~~&~~~\mathcal{A}_p + \mathcal{A}_q \to \mre_{7+p} + \mre_{7+q}\,, ~~~ p,q = 0,1\,,\\
	m = 5,6:~~~&~~~\emptyset \to \mre_8 + \mre_8\,,
	\end{split}
\end{equation}
where we have defined
\begin{equation}\label{newname}
	\mrc_1 \equiv \mra_1\,, ~ \mrf_2 \equiv \mra_2\,, ~ \mrf_3 \equiv \mrb_3%\,, ~ \mre_5 \equiv \mrd_5\,,
	\,, ~ \mrg_1 \equiv \mra_1\,,
        %~ \mathcal{A}_1 \equiv \mra_1\,,
\end{equation}
with the RHS algebras always at level 1. Likewise, the $\mra_1$'s resulting from freezing the $\mre_8$'s in the $\mathbb{Z}_4$-triple are at level 1 (cf. Section \ref{ss:z4}).
%For $m=4$, we denote the $SU(2)$ algebra as $\mathcal{A}_1$ 

%We note that for $m = 4$, the $\mra_1$'s are not uniquely embedded in the momentum lattice $\rii_{3,3}\oplus \mra_1 \oplus \mra_1$. The ones appearing in the above formulas correspond to those explicitly shown in this lattice, while those embedded in $\rii_{3,3}$ remain unaffected upon unfreezing. However, all of them are realized at level 4. 

The converse rules agree perfectly with the freezing mechanism in M-theory on K3 \cite{deBoer:2001wca,Tachikawa:2015wka}. When applied to the enhancements found in the Narain moduli space one reproduces the results, at the level of the algebras, obtained with the exploration algorithm applied to the remaining momentum lattices, as expected.   

\subsection{Full projection map}
\label{ss:proj}
It was shown in \cite{Cvetic:2021sjm} that the list of gauge groups found in the heterotic string on $T^2$, together with their fundamental groups, can be projected to that of the 8d CHL string, generalizing the freezing rules for the algebras discussed above. Namely, consider a gauge group, obtained from the Narain lattice, of the form 
\begin{equation}
	G = \tilde G/H =  G_1 \times \cdots \times G_s \times \sping(2n+16)/H\,,
\end{equation}
where $H$ is generated by an element $k = (k_1,...,k_s, \hat k)$ of the center $Z(\tilde G)$. The corresponding group in the CHL string will be of the form 
\begin{equation}
	G' = G_1 \times \cdots \times G_s \times \spg(n)/H'\,,
\end{equation}
with $H'$ generated by the element $k' = (k_1,...,k_s,\hat k')$ of the center $Z(\tilde G')$. As can be expected, only the contribution of the partially frozen factor will change. Indeed the center of $\sping(2n+16)$ and that of $\spg(n)$ are different. For $n$ odd, we have $\hat k \in \mathbb{Z}_4$ and $\hat k' \in \mathbb{Z}_2$, and the projection reads
\begin{equation}
	\hat k \to \hat k' = \hat k \mod 2 ~~~~~ \left(\{0,1,2,3\} \to \{0,1,0,1\}\right)\,, ~~~n = \text{odd}\,.
\end{equation} 
For $n$ even, we have $\hat k \equiv (\hat k^{(1)},\hat k^{(2)}) \in \mathbb{Z}_2 \times \mathbb{Z}_2$ and again $\hat k' \in \mathbb{Z}_2$, and the projection reads
\begin{equation}
	\hat k \to \hat k' = \hat k^{(1)}+\hat k^{(2)} \mod 2 ~~~~~ \left(\{0,\mathrm{s},\mathrm{c},\mathrm{v}\} \to \{0,1,1,0\}\right)\,, ~~~n = \text{even}\,,
\end{equation} 
where $\{0,\mathrm{s},\mathrm{c},\mathrm{v}\} \equiv \{(0,0),(1,0),(0,1),(1,1)\}$. As a simple example, the gauge group $\sping(32)/\mathbb{Z}_2$ is mapped to $\spg(8)/\mathbb{Z}_2$ \cite{Cvetic:2021sjm,Font:2021uyw}, since the quotient of the former corresponds to a spinor class in the center. In the case that $n = 0$, we lose a simple factor and $(k_1,...,k_s,\hat k)$ goes to $(k_1,...,k_s)$.

This map can be directly generalized to all the different components in the moduli space of 7d theories treated here. Similarly, only the contributions to the fundamental group coming from the partially frozen factors change. In the 7d CHL string the rules for going from $\mrd_{n+4}$ to $\mrc_{n}$ are equivalent to those for going from $\mrd_{n+8}$ to $\mrc_{n}$
described above. For example, we find that $(\sping(24)/\mathbb{Z}_2)\times \sping(14)$ maps to $(\spg(8)/\mathbb{Z}_2) \times \spg(2)$. For the freezing $\mre_{4+n} \to \mrf_n$ (cf. \eqref{frules}), the center of the gauge group is unaltered and so is the corresponding contribution to the fundamental group, i.e. $\hat k \to \hat k' = \hat k$. This is also true for the freezing $\mre_{6+n} \to \mrg_n$ in the $m = 3$ case. 

%For $m = 4$, however, the only partial freezing $\mre_8 \to \mra_1$ does present a change in the center of the gauge group. Our results indicate that the contribution to the fundamental group changes from $\hat k = 0$ to $\hat k' = 1$. In other words, $\mre_8 \to \sug(2)/\mathbb{Z}_2 \simeq \sog(3)$. This can be seen from the fact that the coroot lattice of the rightmost $\mra_1$'s in $\Lambda_4 = \rii_{3,3} \oplus \mra_1 \oplus \mra_1$ embeds into $\Lambda_4^* = \rii_{3,3} \oplus \mra_1^* \oplus \mra_1^*$ such that its real span contains the fundamental weights of each $\mra_1$. 

For $m = 5,6$, the rule $\mre_8 \to \emptyset$ has no effect on the fundamental group other than shortening $(k_1,...,k_s,\hat k)$ to $(k_1,...,k_s)$. With these generalized freezing rules, one can project the enhancements in the Narain component of the moduli space to the other five components treated in this paper to reproduce the results found with our exploration algorithm.

\section{Classification of gauge groups}
\label{s:results}
Now we present the main results of this work and expand in the methods used to obtain them. The full tables with maximal enhancements and their global data are given in Appendix \ref{app:tab}. Here we give tables with the counting of the different gauge symmetries which are realized in each component.

\subsection{Narain Component}
\label{ss:resnar}

Obtaining the gauge groups for the Narain component is done with a straightforward extension of the original exploration algorithm developed in \cite{Font:2020rsk}. Here we have however also computed the complete global data for each group, giving the explicit generators for the fundamental groups using the methods of \cite{Font:2021uyw} based on \cite{Cvetic:2021sjm}. We have for example the gauge group (\# 421 of Table \ref{tab:algebrasT3})
\begin{equation}
	\frac{\sug(8) \times \sug(8) \times \sping(10)}{\mathbb{Z}_8},
\end{equation}
where the fundamental group $\mathbb{Z}_8$ is generated by the element $(1,1,3)$ of the center $\mathbb{Z}_8 \times \mathbb{Z}_8 \times \mathbb{Z}_4$ of the universal cover $\sug(8) \times \sug(8) \times \sping(10)$.

All the maximally enhanced groups in this component are listed in Table \ref{tab:algebrasT3} in Appendix \ref{app:tab1}. The data includes the ADE type of the gauge group and the corresponding fundamental group. The generators of the fundamental group are listed in Table \ref{tab:groupsT3} in Appendix \ref{app:tab2}. For each generator we give a sequence of numbers representing the contribution from the center of each simple factor. In the example just given, the generator is 113. Note that the ordering of the sequence corresponds to the ordering of the listed ADE type. To properly read the sequence one must write expressions of the form $\mra_3^2 \mrd_4^3$ as $(\mra_3,\mra_3,\mrd_4,\mrd_4,\mrd_4)$, e.g, assigning each number in the sequence to each subsequent ADE factor. For $\mrd_{2n}$ factors there are four order two elements in the center denoted v, c, s and 1, corresponding to the vector class, spinor classes and the identity, respectively. Note that in some cases the fundamental group has more than one generator.

The total number of distinct gauge algebras and distinct gauge groups for different ranks of the semisimple part are listed in Table \ref{tab:numbersT3}. These have been obtained by deleting nodes in the Dynkin Diagrams of the maximally enhanced groups, and we assume that this gives all the possibilities, as discussed in Section \ref{sss:alg}.\\

{\centering\scriptsize
	\setlength{\tabcolsep}{2.5pt}%
	\begin{tabular}
		{
			|
			>{$}c<{$}|
			>{$}c<{$}
			>{$}c<{$}
			>{$}c<{$}
			>{$}c<{$}
			>{$}c<{$}
			>{$}c<{$}
			>{$}c<{$}
			>{$}c<{$}
			>{$}c<{$}
			>{$}c<{$}
			>{$}c<{$}
			>{$}c<{$}
			>{$}c<{$}
			>{$}c<{$}
			>{$}c<{$}
			>{$}c<{$}|
			>{$}c<{$}
			>{$}c<{$}|}
		\hline
		\text{Rank} & 1 & \mathbb{Z}_2 & \mathbb{Z}_2{}^2 & \mathbb{Z}_3 & \mathbb{Z}_4 & \mathbb{Z}_2{}^3 & \mathbb{Z}_2{}^4 &\mathbb{Z}_2{}^5& \mathbb{Z}_5 & \mathbb{Z}_6 & \mathbb{Z}_3{}^2 & \mathbb{Z}_2 \mathbb{Z}_4 & \mathbb{Z}_7 & \mathbb{Z}_2 \mathbb{Z}_6 & \mathbb{Z}_4{}^2 & \mathbb{Z}_8 & \text{Algebras} & \text{Groups} \\
		\hline
		19 & 652 & 381 & 68 & 51 & 37 & 5 & 1 & & 6 & 16 & 3 & 2 & 1 & 2 & 1 & 2 & 1035 & 1232 \\
		\hline
		18 & 852 & 492 & 89 & 52 & 35 & 9 & 1 & & 4 & 10 & 3 & 6 & 1 & 1 & 1 & 1 & 1180 & 1557 \\
		\hline
		17 & 827 & 442 & 73 & 39 & 23 & 8 & 1 && 2 & 4 & 2 & 3 & \text{} & \text{} & \text{} & \text{} & 1024 & 1424 \\
		\hline
		16 & 694 & 334 & 47 & 25 & 12 & 4 & 1 &1& 1 & 1 & 1 & 1 & \text{} & \text{} & \text{} & \text{} & 794 & 1122 \\
		\hline
		15 & 528 & 217 & 24 & 12 & 4 & 2 & 1 && \text{} & \text{} & \text{} & \text{} & \text{} & \text{} & \text{} & \text{} & 567 & 788 \\
		\hline
		14 & 389 & 128 & 11 & 6 & 1 & 1 & \text{} & \text{} & \text{} & \text{} & \text{} & \text{} & \text{} & &\text{} & \text{} & 403 & 536 \\
		\hline
		13 & 272 & 66 & 3 & 2 & \text{} & \text{} & \text{} & \text{} & \text{} & \text{} & \text{} & \text{} && \text{} & \text{} & \text{} & 276 & 343 \\
		\hline
		12 & 192 & 33 & 1 & 1 & \text{} & \text{} & \text{} & \text{} & \text{} && \text{} & \text{} & \text{} & \text{} & \text{} & \text{} & 193 & 227 \\
		\hline
		11 & 128 & 14 & \text{} & \text{} & \text{} & \text{} & \text{} & &\text{} & \text{} & \text{} & \text{} & \text{} & \text{} & \text{} & \text{} & 128 & 142 \\
		\hline
		10 & 88 & 6 & \text{} & \text{} & \text{} && \text{} & \text{} & \text{} & \text{} & \text{} & \text{} & \text{} & \text{} & \text{} & \text{} & 88 & 94 \\
		\hline
		9 & 57 & 2 & \text{} & \text{} & \text{} && \text{} & \text{} & \text{} & \text{} & \text{} & \text{} & \text{} & \text{} & \text{} & \text{} & 57 & 59 \\
		\hline
		8 & 39 & 1 & \text{} & \text{} & \text{} && \text{} & \text{} & \text{} & \text{} & \text{} & \text{} & \text{} & \text{} & \text{} & \text{} & 39 & 40 \\
		\hline
		7 & 24 & \text{} & \text{} & \text{} && \text{} & \text{} & \text{} & \text{} & \text{} & \text{} & \text{} & \text{} & \text{} & \text{} & \text{} & 24 & 24 \\
		\hline
		6 & 16 & \text{} & \text{} & \text{} & \text{} & \text{} & \text{} && \text{} & \text{} & \text{} & \text{} & \text{} & \text{} & \text{} & \text{} & 16 & 16 \\
		\hline
		5 & 9 & \text{} & \text{} & \text{} & \text{} & \text{} & \text{} && \text{} & \text{} & \text{} & \text{} & \text{} & \text{} & \text{} & \text{} & 9 & 9 \\
		\hline
		4 & 6 & \text{} & \text{} & \text{} & \text{} & \text{} & \text{} & &\text{} & \text{} & \text{} & \text{} & \text{} & \text{} & \text{} & \text{} & 6 & 6 \\
		\hline
		3 & 3 & \text{} & \text{} & \text{} & \text{} & \text{} & \text{} & &\text{} & \text{} & \text{} & \text{} & \text{} & \text{} & \text{} & \text{} & 3 & 3 \\
		\hline
		2 & 2 & \text{} & \text{} & \text{} & \text{} & \text{} & \text{} && \text{} & \text{} & \text{} & \text{} & \text{} & \text{} & \text{} & \text{} & 2 & 2 \\
		\hline
		1 & 1 & \text{} & \text{} & \text{} & \text{} & \text{} & \text{} & &\text{} & \text{} & \text{} & \text{} & \text{} & \text{} & \text{} & \text{} & 1 & 1 \\
		\hline
		\text{All} & 4779 & 2116 & 316 & 188 & 112 & 29 & 5 & 1&13 & 31 & 9 & 12 & 2 & 3 & 2 & 3 & 5845 & 7625 \\
		\hline
	\end{tabular}
	\begin{table}[H]
		\caption{Number of algebras and groups of each rank with a certain fundamental group for the heterotic string on $T^3$. The gauge group with $\pi_1 = \mathbb{Z}_2^5$ (cf. eq. \eqref{noenh}) does not admit further enhancements. }
		\label{tab:numbersT3}
	\end{table}
}
We note that there are many cases in which two gauge groups have isomorphic fundamental groups with inequivalent inclusions in the center of the universal covering (meaning that they are not related by outer automorphisms of the group, as is the case e.g. for $\sog(2n)$ versus $\sping(2n)/\mathbb{Z}_2$ for $n \neq 4$). These are not distinguished in Table \ref{tab:algebrasT3}, so that the numbering goes only up to 1163. The inequivalence is taken into account in Table \ref{tab:groupsT3} by putting primes on the corresponding numbering. 

\subsection{Triples}
\label{ss:restrip}
The results for the components of the moduli space with rank reduction are obtained by an extension of the exploration algorithm taking into account Proposition \ref{propTrip}. The gauge groups are recorded in Tables \ref{tab:algebrasZ2} to \ref{tab:algebrasZ5and6} in Appendix \ref{app:tab1}, while the generators for the fundamental groups are recorded in Tables \ref{tab:groupsZ2} and \ref{tab:groupsZ3} in Appendix \ref{app:tab2}. In the case of the $\mathbb{Z}_5$ and $\mathbb{Z}_6$-triples all of the gauge groups are simply connected and so no global data is required to specify them. The data is presented with the same conventions as for the Narain component, together with the notation defined in eq. \eqref{newname}. As explained in Section \ref{ss:proj}, all the gauge groups for the non-trivial $\mathbb{Z}_m$ triples can be obtained from those of the Narain component using a projection map generalizing the one obtained in \cite{Cvetic:2021sjm} for the 8d CHL string. The total number of distinct gauge algebras and distinct gauge groups are listed in Table \ref{tab:numbersTriples}.\\

{\centering\scriptsize
	\setlength{\tabcolsep}{1.5pt}%
	\begin{tabular}[b]
		{
			|
			>{$}c<{$}|
			>{$}c<{$}
			>{$}c<{$}
			>{$}c<{$}
			>{$}c<{$}
			>{$}c<{$}
			>{$}c<{$}
			>{$}c<{$}
			>{$}c<{$}
			>{$}c<{$}|}
		\multicolumn{10}{c}{\small$\mathbb{Z}_2$ triple}\\
\hline
 \text{Rank} & 1 & \mathbb{Z}_2 & \mathbb{Z}_2{}^2 & \mathbb{Z}_3 & \mathbb{Z}_4 & \mathbb{Z}_2{}^3 & \mathbb{Z}_2{}^4 & \text{Algebras} & \text{Groups} \\
\hline
 11 & 224 & 143 & 44 & 7 & 3 & 7 & 1 & 407 & 429 \\
\hline
 10 & 307 & 192 & 51 & 5 & 3 & 8 & 1 & 473 & 567 \\
\hline
 9 & 284 & 161 & 37 & 2 & 2 & 4 & 1 & 372 & 491 \\
\hline
 8 & 214 & 101 & 18 & 1 & 1 & 2 & \text{} & 244 & 337 \\
\hline
 7 & 137 & 45 & 5 & \text{} & \text{} & \text{} & \text{} & 143 & 187 \\
\hline
 6 & 84 & 17 & 1 & \text{} & \text{} & \text{} & \text{} & 85 & 102 \\
\hline
 5 & 46 & 4 & \text{} & \text{} & \text{} & \text{} & \text{} & 46 & 50 \\
\hline
 4 & 26 & 1 & \text{} & \text{} & \text{} & \text{} & \text{} & 26 & 27 \\
\hline
 3 & 12 & \text{} & \text{} & \text{} & \text{} & \text{} & \text{} & 12 & 12 \\
\hline
 2 & 6 & \text{} & \text{} & \text{} & \text{} & \text{} & \text{} & 6 & 6 \\
\hline
 1 & 2 & \text{} & \text{} & \text{} & \text{} & \text{} & \text{} & 2 & 2 \\
\hline
 \text{All} & 1342 & 664 & 156 & 15 & 9 & 21 & 3 & 1816 & 2210 \\
\hline
\end{tabular}
\begin{tabular}[b]
		{
			|
			>{$}c<{$}|
			>{$}c<{$}
			>{$}c<{$}
			>{$}c<{$}
			>{$}c<{$}
			>{$}c<{$}|}
		\multicolumn{6}{c}{\small$\mathbb{Z}_3$ triple}\\
\hline
 \text{Rank} & 1 & \mathbb{Z}_2 & \mathbb{Z}_3 & \text{Algebras} & \text{Groups} \\
\hline
 7 & 41 & 6 & 5 & 50 & 52 \\
\hline
 6 & 37 & 5 & 4 & 41 & 46 \\
\hline
 5 & 24 & 2 & 2 & 24 & 28 \\
\hline
 4 & 15 & 1 & 1 & 15 & 17 \\
\hline
 3 & 8 & \text{} & \text{} & 8 & 8 \\
\hline
 2 & 5 & \text{} & \text{} & 5 & 5 \\
\hline
 1 & 2 & \text{} & \text{} & 2 & 2 \\
\hline
 \text{All} & 132 & 14 & 12 & 145 & 158 \\
\hline 
\multicolumn{6}{c}{}\\
\multicolumn{6}{c}{}\\
\end{tabular}
\begin{tabular}[b]
		{
			|
			>{$}c<{$}|
			>{$}c<{$}
			>{$}c<{$}
			>{$}c<{$}
			>{$}c<{$}|}
		\multicolumn{5}{c}{\small$\mathbb{Z}_4$ triple}\\
		\hline
		\text{Rank} & 1 & \mathbb{Z}_2 & \text{Algebras} & \text{Groups} \\
		\hline
		5 & 13 & 5 & 16 & 18 \\\hline
		4 & 10 & 4 & 11 & 14 \\\hline
		3 & 6 & 2 & 6 & 8 \\\hline
		2 & 4 & 1 & 4 & 5 \\\hline
		1 & 2 & \text{} & 2 & 2 \\\hline
		\text{All} & 35 & 12 & 39 & 47 \\
		\hline
		\multicolumn{5}{c}{\small$\mathbb{Z}_5$ and $\mathbb{Z}_6$ triples}\\
		\hline
		\text{Rank} & 1 && \text{Algebras} & \text{Groups} \\
		\hline
		3 & 3 && 3 & 3 \\
		\hline
		2 & 2 && 2 & 2 \\
		\hline
		1 & 1 && 1 & 1 \\
		\hline
		\text{All} & 6 && 6 & 6 \\
		\hline
	\end{tabular}
	\begin{table}[H]
		\caption{Number of algebras and groups of each rank with a certain fundamental group for the heterotic $\mathbb{Z}_2$, $\mathbb{Z}_3$, $\mathbb{Z}_4$, $\mathbb{Z}_5$ and $\mathbb{Z}_6$ triples.}
		\label{tab:numbersTriples}
	\end{table}
}

\section{Conclusions}
In this paper we have advanced the classification of possible gauge groups realized in the string landscape to three-dimensional internal manifolds when the number of supercharges is 16. This was done by finding embeddings of weight lattices into the momentum lattices constructed in \cite{deBoer:2001wca}, taking into account an extra constraint on the role of the lattice vectors as stated in Proposition \ref{propTrip}. We have however ignored other components with further rank reduction (see e.g. \cite{Aharony:2007du,Dabholkar:1996pc,Montero:2020icj}), which to our knowledge have so far no heterotic description. 

We have also studied the mechanism of partial singularity freezing found in M-theory on K3 from the heterotic side, using lattice embedding techniques. We have also constructed the more general freezing rules which indicate how the topology of the gauge group changes, generalizing the results of \cite{Cvetic:2021sjm} for the 8d CHL string. It is clear that this mechanism can be generalized in the heterotic string to compactifications to lower dimensions, e.g. on $T^4$. We plan on investigating this point further in a future work.

Finally, we note that our results may serve to test Swampland conjectures \cite{Vafa:2005ui}, which are also easier to study in high dimensional theories with high supersymmetry (see e.g. \cite{Montero:2020icj,Hamada:2021bbz,Cvetic:2020kuw}). More generally it would be interesting to determine with more confidence if there are or not more components in the moduli space of heterotic strings with 16 supercharges using the techniques of asymmetric orbifolds. Recent results \cite{Montero:2020icj,Hamada:2021bbz} indicate that this is not the case in 9d and 8d, but the 7d case remains in question. 

\subsection*{Acknowledgements}
We are grateful to Ruben Minasian and Miguel Montero for interesting and stimulating conversations, and Mariana Gra\~na for valuable guidance and supervision in the development of this project. We also thank Anamaría Font and Carmen Nuñez for their collaboration in initiating and developing this research direction in previous works.
This work was partially supported
by the ERC Consolidator Grant 772408-Stringlandscape, PIP-CONICET-11220150100559CO, UBACyT 2018-2021 and ANPCyT-PICT-2016-1358 (2017-2020).

\newpage
\appendix

\section{Comments on lattices technicalities}
\label{app}
In this appendix we discuss some technicalities regarding lattice embeddings in the context of the heterotic string and its compactifications considered in the paper, namely the concept of a primitive embedding and of lattice isomorphisms.

\subsection{Primitive embeddings}
\label{app:prim}
The Narain moduli space is usually understood as the space of embeddings of the even self-dual Lorentzian lattice $\rii_{16+d,d}$ into an ambient space $\mathbb{R}^{16+d,d}$. The roots of the gauge group are associated to lattice vectors $(P,p_L;p_R)$ subject to the constraints $P^2 + p_L^2 = 2$ and $p_R = 0$, hence they are precisely the vectors of squared norm 2 which lie in the positive definite subspace $\mathbb{R}^{16+d}$. When one rotates and/or boosts the lattice by an $SO(d+16,d)$ transformation, one moves through the moduli space and the set of vectors intersecting $\mathbb{R}^{16+d}$ changes. This corresponds to changes in the spectrum, particularly in the gauge symmetry group. 

Conversely we may think of this problem as fixing the Narain lattice itself and transforming instead the ambient space. Intuitively we can think of rotating the positive-definite $(16+d)$-plane in relation to the lattice, whose origin it intersects. This intuition can be formalized by noting that the moduli space is indeed (locally) a Grassmann space:
\begin{equation}
\mathcal{M} = \sog(16+d,d)\big/\sog(16+d)\times \sog(d) = \text{Gr}(16+d,d).
\end{equation}
Each model in moduli space corresponds therefore to a choice of a $(16+d)$-plane whose intersection with the Narain lattice yields the data for the gauge group. Indeed, this intersection contains vectors which are not necessarily constrained to have squared norm 2, which in the case of not being sums of roots will correspond to weights. Because of this we refer to the intersecting lattice as $M$. 

The weight lattice $M$ is an example of a primitive embedding. Technically, a primitive embedding of a lattice $L$ into another lattice $M$ is such that the intersection of the real span of $L$ with $M$ is $L$ itself. A non-primitive embedding is provided by any proper sublattice $M'$ of $M$ of the same rank, since its real span will contain vectors of $M$ not in $M'$. An example of this is the embedding of the lattice $\mrd_8$ into $\mre_8$, as the real span of $\mrd_8$ contains the spinor weight which extends it to $\mre_8$.

From this discussion we see that the natural way of thinking about sublattices of the Narain lattice (ant other momentum lattices) is by considering primitive embeddings, which are in general not root lattices, but \textit{weight} lattices. There are of course other possibilities such as lattices with no roots and hence no weights, but we do not consider these cases as they play no role in the classification of the gauge algebra performed here.

\subsection{Lattice Isomorphisms}
\label{app:iso}
In this paper we have chosen to describe the momentum lattices in ways that naturally reflect the gauge groups that they yield in the respective string theory. The simplest example is that of the Mikhailov lattice in 8d, which was originally written as
\begin{equation}
\rii_{1,1}(2) \oplus \rii_{1,1} \oplus \mre_8 \simeq \rii_{2,2} \oplus  \mrd_8.
\end{equation} 
We write it instead as
\begin{equation}
\rii_{2,2} \oplus \mrc_8,
\end{equation}
where the lattice isomorphism $\mrd_n \simeq \mrc_n$ has been employed. Indeed, both lattices are exactly the same, but the \textit{algebras} are not. For instance, one finds the lattice $\mrc_{8}$ in the Narain moduli space, but its long root does not give a massless state and so the gauge algebra is $\mrd_{8}$. In the CHL string however one can obtain such a massless state and indeed realize the gauge algebra $\mrc_8$.

For the CHL string in 7d we similarly use the lattice isomorphism $\mrd_4 \simeq \mrf_4$, and the story is similar to the 8d case just discussed.  This extends to the $\mathbb{Z}_3$-triple where one finds the gauge algebra $2 \mrg_2$ at the canonical point in moduli space and not $2 \mra_2$. 
\vspace{2em}
\section{Maximal enhancements for 7d heterotic string}
\label{app:tab}
In this appendix we record the maximally enhanced gauge groups realized in the 7d $\mathbb{Z}_m$-triples constructed from the heterotic string. The algebras and the fundamental groups are presented in Appendix \ref{app:tab1}, while the generators of the fundamental group are presented in Appendix \ref{app:tab2}. The way in which the data is encoded is explained in Section \ref{s:results}.\\
\subsection{Maximally enhanced algebras and fundamental groups}\label{app:tab1}
\newpage
\renewcommand{\baselinestretch}{1.5}
{\centering\scriptsize
	\setlength{\tabcolsep}{1pt}%
	\setlength\arrayrulewidth{1pt}%
\begin{tabular}
		{
			|>{\columncolor{Gray}$}c<{$}>{\columncolor{white}$}l<{$}>{\columncolor{white}$}c<{$}
			|>{\columncolor{Gray}$}c<{$}>{\columncolor{white}$}l<{$}>{\columncolor{white}$}c<{$}
			|>{\columncolor{Gray}$}c<{$}>{\columncolor{white}$}l<{$}>{\columncolor{white}$}c<{$}
			|>{\columncolor{Gray}$}c<{$}>{\columncolor{white}$}l<{$}>{\columncolor{white}$}c<{$}
			|>{\columncolor{Gray}$}c<{$}>{\columncolor{white}$}l<{$}>{\columncolor{white}$}c<{$}
			|>{\columncolor{Gray}$}c<{$}>{\columncolor{white}$}l<{$}>{\columncolor{white}$}c<{$}
			|
		}\hline	
		1 & \text{A}_1 \text{A}_3^6 & \mathbb{Z}_4^2 & 48 & \text{A}_2^4 \text{A}_5 \text{A}_6 & \mathbb{Z}_3 & 95 & \text{A}_1^2 \text{A}_5^2 \text{A}_7 & \mathbb{Z}_2 & 142 & \text{A}_1^2 \text{A}_4 \text{A}_5 \text{A}_8 & \text{} & 189 & \text{A}_1^4 \text{A}_6 \text{A}_9 & \mathbb{Z}_2 \\
		2 & \text{A}_3^5 \text{A}_4 & \mathbb{Z}_4 & 49 & \text{A}_1 \text{A}_2^2 \text{A}_3 \text{A}_5 \text{A}_6 & \text{} & 96 & \text{A}_1^2 \text{A}_5^2 \text{A}_7 & \mathbb{Z}_2^2 & 143 & \text{A}_2 \text{A}_4 \text{A}_5 \text{A}_8 & \text{} & 190 & \text{A}_1^2 \text{A}_2 \text{A}_6 \text{A}_9 & \text{} \\
		3 & \text{A}_1 \text{A}_2^2 \text{A}_3^2 \text{A}_4^2 & \text{} & 50 & \text{A}_1^2 \text{A}_3^2 \text{A}_5 \text{A}_6 & \mathbb{Z}_2 & 97 & \text{A}_2 \text{A}_5^2 \text{A}_7 & \text{} & 144 & \text{A}_2 \text{A}_4 \text{A}_5 \text{A}_8 & \mathbb{Z}_3 & 191 & \text{A}_2^2 \text{A}_6 \text{A}_9 & \text{} \\
		4 & \text{A}_1^2 \text{A}_3^3 \text{A}_4^2 & \mathbb{Z}_2 & 51 & \text{A}_2 \text{A}_3^2 \text{A}_5 \text{A}_6 & \text{} & 98 & \text{A}_1^2 \text{A}_2^2 \text{A}_6 \text{A}_7 & \text{} & 145 & \text{A}_1 \text{A}_5^2 \text{A}_8 & \text{} & 192 & \text{A}_1 \text{A}_3 \text{A}_6 \text{A}_9 & \text{} \\
\hline
	\end{tabular}
	
	\begin{tabular}
		{
			|>{\columncolor{Gray}$}c<{$}>{\columncolor{white}$}l<{$}>{\columncolor{white}$}c<{$}
			|>{\columncolor{Gray}$}c<{$}>{\columncolor{white}$}l<{$}>{\columncolor{white}$}c<{$}
			|>{\columncolor{Gray}$}c<{$}>{\columncolor{white}$}l<{$}>{\columncolor{white}$}c<{$}
			|>{\columncolor{Gray}$}c<{$}>{\columncolor{white}$}l<{$}>{\columncolor{white}$}c<{$}
			|>{\columncolor{Gray}$}c<{$}>{\columncolor{white}$}l<{$}>{\columncolor{white}$}c<{$}
			|>{\columncolor{Gray}$}c<{$}>{\columncolor{white}$}l<{$}>{\columncolor{white}$}c<{$}
			|
		}\hline	
		236 & \text{A}_2 \text{A}_3^2 \text{A}_{11} & \mathbb{Z}_4 & 283 & \text{A}_6 \text{A}_{13} & \text{} & 330 & \text{A}_2 \text{A}_3 \text{A}_4 \text{A}_6 \text{D}_4 & \text{} & 377 & \text{A}_2 \text{A}_3^3 \text{D}_4^2 & \mathbb{Z}_2^2 & 424 & \text{A}_1^2 \text{A}_4 \text{A}_8 \text{D}_5 & \text{} \\
		237 & \text{A}_1^4 \text{A}_4 \text{A}_{11} & \mathbb{Z}_2 & 284 & \text{A}_1^3 \text{A}_2 \text{A}_{14} & \text{} & 331 & \text{A}_1 \text{A}_4^2 \text{A}_6 \text{D}_4 & \text{} & 378 & \text{A}_1 \text{A}_5^2 \text{D}_4^2 & \mathbb{Z}_2^2 & 425 & \text{A}_2 \text{A}_4 \text{A}_8 \text{D}_5 & \text{} \\
		238 & \text{A}_1^2 \text{A}_2 \text{A}_4 \text{A}_{11} & \text{} & 285 & \text{A}_1^3 \text{A}_2 \text{A}_{14} & \mathbb{Z}_3 & 332 & \text{A}_1 \text{A}_3 \text{A}_5 \text{A}_6 \text{D}_4 & \mathbb{Z}_2 & 379 & \text{A}_2^2 \text{A}_7 \text{D}_4^2 & \mathbb{Z}_2 & 426 & \text{A}_1 \text{A}_5 \text{A}_8 \text{D}_5 & \text{} \\
		239 & \text{A}_1^2 \text{A}_2 \text{A}_4 \text{A}_{11} & \mathbb{Z}_2 & 286 & \text{A}_1 \text{A}_2^2 \text{A}_{14} & \text{} & 333 & \text{A}_4 \text{A}_5 \text{A}_6 \text{D}_4 & \text{} & 380 & \text{A}_1^3 \text{D}_4^4 & \mathbb{Z}_2^4 & 427 & \text{A}_6 \text{A}_8 \text{D}_5 & \text{} \\
\hline
	\end{tabular}
	
	\begin{tabular}
		{
			|>{\columncolor{Gray}$}c<{$}>{\columncolor{white}$}l<{$}>{\columncolor{white}$}c<{$}
			|>{\columncolor{Gray}$}c<{$}>{\columncolor{white}$}l<{$}>{\columncolor{white}$}c<{$}
			|>{\columncolor{Gray}$}c<{$}>{\columncolor{white}$}l<{$}>{\columncolor{white}$}c<{$}
			|>{\columncolor{Gray}$}c<{$}>{\columncolor{white}$}l<{$}>{\columncolor{white}$}c<{$}
			|>{\columncolor{Gray}$}c<{$}>{\columncolor{white}$}l<{$}>{\columncolor{white}$}c<{$}
			|>{\columncolor{Gray}$}c<{$}>{\columncolor{white}$}l<{$}>{\columncolor{white}$}c<{$}
			|
		}\hline	
		471 & \text{A}_1^2 \text{A}_7 \text{D}_5^2 & \mathbb{Z}_4 & 518 & \text{A}_5 \text{A}_8 \text{D}_6 & \text{} & 565 & \text{A}_1 \text{D}_6^3 & \mathbb{Z}_2^2 & 612 & \text{A}_2^2 \text{A}_4 \text{D}_4 \text{D}_7 & \text{} & 659 & \text{A}_2^2 \text{A}_7 \text{D}_8 & \mathbb{Z}_2 \\
		472 & \text{A}_2 \text{A}_7 \text{D}_5^2 & \mathbb{Z}_4 & 519 & \text{A}_1^2 \text{A}_2 \text{A}_9 \text{D}_6 & \mathbb{Z}_2 & 566 & \text{A}_1 \text{A}_2 \text{A}_3^3 \text{D}_7 & \mathbb{Z}_4 & 613 & \text{A}_1 \text{A}_2 \text{A}_5 \text{D}_4 \text{D}_7 & \mathbb{Z}_2 & 660 & \text{A}_1 \text{A}_3 \text{A}_7 \text{D}_8 & \mathbb{Z}_2 \\
		473 & \text{A}_1 \text{A}_8 \text{D}_5^2 & \text{} & 520 & \text{A}_2^2 \text{A}_9 \text{D}_6 & \text{} & 567 & \text{A}_3^4 \text{D}_7 & \mathbb{Z}_4 & 614 & \text{A}_2 \text{A}_6 \text{D}_4 \text{D}_7 & \text{} & 661 & \text{A}_4 \text{A}_7 \text{D}_8 & \mathbb{Z}_2 \\
		474 & \text{A}_9 \text{D}_5^2 & \text{} & 521 & \text{A}_2^2 \text{A}_9 \text{D}_6 & \mathbb{Z}_2 & 568 & \text{A}_1^2 \text{A}_2^3 \text{A}_4 \text{D}_7 & \text{} & 615 & \text{A}_1 \text{A}_7 \text{D}_4 \text{D}_7 & \mathbb{Z}_2 & 662 & \text{A}_1 \text{A}_2 \text{A}_8 \text{D}_8 & \text{} \\
\hline
	\end{tabular}
	
	\begin{tabular}
		{
			|>{\columncolor{Gray}$}c<{$}>{\columncolor{white}$}l<{$}>{\columncolor{white}$}c<{$}
			|>{\columncolor{Gray}$}c<{$}>{\columncolor{white}$}l<{$}>{\columncolor{white}$}c<{$}
			|>{\columncolor{Gray}$}c<{$}>{\columncolor{white}$}l<{$}>{\columncolor{white}$}c<{$}
			|>{\columncolor{Gray}$}c<{$}>{\columncolor{white}$}l<{$}>{\columncolor{white}$}c<{$}
			|>{\columncolor{Gray}$}c<{$}>{\columncolor{white}$}l<{$}>{\columncolor{white}$}c<{$}
			|>{\columncolor{Gray}$}c<{$}>{\columncolor{white}$}l<{$}>{\columncolor{white}$}c<{$}
			|
		}\hline	
		706 & \text{A}_2^2 \text{A}_6 \text{D}_9 & \text{} & 753 & \text{A}_1 \text{A}_4 \text{D}_4 \text{D}_{10} & \mathbb{Z}_2 & 800 & \text{D}_7 \text{D}_{12} & \mathbb{Z}_2 & 847 & \text{A}_3 \text{A}_5^2 \text{E}_6 & \text{} & 894 & \text{A}_1 \text{A}_4^2 \text{D}_4 \text{E}_6 & \text{} \\
		707 & \text{A}_1 \text{A}_3 \text{A}_6 \text{D}_9 & \text{} & 754 & \text{A}_5 \text{D}_4 \text{D}_{10} & \mathbb{Z}_2 & 801 & \text{A}_1^2 \text{A}_2^2 \text{D}_{13} & \text{} & 848 & \text{A}_3 \text{A}_5^2 \text{E}_6 & \mathbb{Z}_2 & 895 & \text{A}_2^2 \text{A}_5 \text{D}_4 \text{E}_6 & \mathbb{Z}_3 \\
		708 & \text{A}_4 \text{A}_6 \text{D}_9 & \text{} & 755 & \text{A}_1^2 \text{A}_2 \text{D}_5 \text{D}_{10} & \mathbb{Z}_2 & 802 & \text{A}_1 \text{A}_2 \text{A}_3 \text{D}_{13} & \text{} & 849 & \text{A}_3 \text{A}_5^2 \text{E}_6 & \mathbb{Z}_3 & 896 & \text{A}_1 \text{A}_3 \text{A}_5 \text{D}_4 \text{E}_6 & \mathbb{Z}_2 \\
		709 & \text{A}_1^3 \text{A}_7 \text{D}_9 & \mathbb{Z}_2 & 756 & \text{A}_2^2 \text{D}_5 \text{D}_{10} & \text{} & 803 & \text{A}_1^2 \text{A}_4 \text{D}_{13} & \text{} & 850 & \text{A}_3 \text{A}_5^2 \text{E}_6 & \mathbb{Z}_6 & 897 & \text{A}_4 \text{A}_5 \text{D}_4 \text{E}_6 & \text{} \\
\hline
	\end{tabular}
	
	\begin{tabular}
		{
			|>{\columncolor{Gray}$}c<{$}>{\columncolor{white}$}l<{$}>{\columncolor{white}$}c<{$}
			|>{\columncolor{Gray}$}c<{$}>{\columncolor{white}$}l<{$}>{\columncolor{white}$}c<{$}
			|>{\columncolor{Gray}$}c<{$}>{\columncolor{white}$}l<{$}>{\columncolor{white}$}c<{$}
			|>{\columncolor{Gray}$}c<{$}>{\columncolor{white}$}l<{$}>{\columncolor{white}$}c<{$}
			|>{\columncolor{Gray}$}c<{$}>{\columncolor{white}$}l<{$}>{\columncolor{white}$}c<{$}
			|>{\columncolor{Gray}$}c<{$}>{\columncolor{white}$}l<{$}>{\columncolor{white}$}c<{$}
			|>{\columncolor{Gray}$}c<{$}>{\columncolor{white}$}l<{$}>{\columncolor{white}$}c<{$}
			|
		}\hline	
		941 & \text{A}_1^3 \text{D}_{10} \text{E}_6 & \mathbb{Z}_2 & 979 & \text{A}_3 \text{A}_4 \text{A}_5 \text{E}_7 & \mathbb{Z}_2 & 1017 & \text{A}_1 \text{A}_2 \text{A}_5 \text{D}_4 \text{E}_7 & \mathbb{Z}_2 & 1055 & \text{D}_5 \text{D}_7 \text{E}_7 & \text{} & 1093 & \text{A}_1^2 \text{A}_2 \text{A}_3 \text{A}_4 \text{E}_8 & \text{} & 1131 & \text{A}_2 \text{A}_4 \text{D}_5 \text{E}_8 & \text{} \\
		942 & \text{A}_1 \text{A}_2 \text{D}_{10} \text{E}_6 & \text{} & 980 & \text{A}_1^2 \text{A}_5^2 \text{E}_7 & \mathbb{Z}_2 & 1018 & \text{A}_3 \text{A}_5 \text{D}_4 \text{E}_7 & \mathbb{Z}_2 & 1056 & \text{A}_1^4 \text{D}_8 \text{E}_7 & \mathbb{Z}_2^2 & 1094 & \text{A}_2^2 \text{A}_3 \text{A}_4 \text{E}_8 & \text{} & 1132 & \text{A}_1 \text{A}_5 \text{D}_5 \text{E}_8 & \text{} \\
		943 & \text{A}_3 \text{D}_{10} \text{E}_6 & \text{} & 981 & \text{A}_2 \text{A}_5^2 \text{E}_7 & \text{} & 1019 & \text{A}_2 \text{A}_6 \text{D}_4 \text{E}_7 & \text{} & 1057 & \text{A}_1^2 \text{A}_2 \text{D}_8 \text{E}_7 & \mathbb{Z}_2 & 1095 & \text{A}_1 \text{A}_3^2 \text{A}_4 \text{E}_8 & \text{} & 1133 & \text{A}_6 \text{D}_5 \text{E}_8 & \text{} \\
\hline
	\end{tabular} \begin{table}[H]
		\caption{Algebras of maximal rank for the heterotic string on $T^3$.}
		\label{tab:algebrasT3}
	\end{table}
}

{\centering\scriptsize
	\setlength{\tabcolsep}{1pt}%
	\setlength\arrayrulewidth{1pt}%
	\begin{tabular}
		{
			|>{\columncolor{Gray}$}c<{$}>{\columncolor{white}$}l<{$}>{\columncolor{white}$}c<{$}
			|>{\columncolor{Gray}$}c<{$}>{\columncolor{white}$}l<{$}>{\columncolor{white}$}c<{$}
			|>{\columncolor{Gray}$}c<{$}>{\columncolor{white}$}l<{$}>{\columncolor{white}$}c<{$}
			|>{\columncolor{Gray}$}c<{$}>{\columncolor{white}$}l<{$}>{\columncolor{white}$}c<{$}
			|>{\columncolor{Gray}$}c<{$}>{\columncolor{white}$}l<{$}>{\columncolor{white}$}c<{$}
			|>{\columncolor{Gray}$}c<{$}>{\columncolor{white}$}l<{$}>{\columncolor{white}$}c<{$}
			|
		}\hline
	 1 & \text{A}_1^2 \text{A}_3^3 & \mathbb{Z}_2^3 & 48 & \text{A}_3 \text{A}_4 \text{C}_2^2 & \mathbb{Z}_2 & 95 & \text{A}_1 \text{A}_5 \text{C}_5 & \mathbb{Z}_2 & 142 & \text{A}_4 \text{C}_1^2 \text{D}_5 & \text{} & 189 & \text{C}_1 \text{C}_3 \text{E}_7 & \text{} & 236 & \text{A}_1^2 \text{C}_7 \text{F}_2 & \text{} \\
 2 & \text{A}_2 \text{A}_3^3 & \mathbb{Z}_2^2 & 49 & \text{A}_1^2 \text{A}_5 \text{C}_2^2 & \mathbb{Z}_2^2 & 96 & \text{A}_1 \text{A}_2^2 \text{C}_1 \text{C}_5 & \text{} & 143 & \text{A}_1 \text{A}_3 \text{C}_2 \text{D}_5 & \mathbb{Z}_2^2 & 190 & \text{A}_1 \text{C}_1^2 \text{E}_8 & \text{} & 237 & \text{A}_2 \text{C}_7 \text{F}_2 & \text{} \\
 3 & \text{A}_1 \text{A}_5^2 & \mathbb{Z}_2^2 & 50 & \text{A}_2 \text{A}_5 \text{C}_2^2 & \mathbb{Z}_2 & 97 & \text{A}_1 \text{A}_4 \text{C}_1 \text{C}_5 & \text{} & 144 & \text{A}_1 \text{A}_2 \text{C}_1 \text{C}_2 \text{D}_5 & \mathbb{Z}_2 & 191 & \text{C}_1 \text{C}_2 \text{E}_8 & \text{} & 238 & \text{A}_1 \text{C}_8 \text{F}_2 & \text{} \\
 4 & \text{A}_2^2 \text{A}_7 & \mathbb{Z}_2 & 51 & \text{A}_7 \text{C}_2^2 & \mathbb{Z}_2 & 98 & \text{A}_5 \text{C}_1 \text{C}_5 & \text{} & 145 & \text{A}_1^2 \text{C}_2^2 \text{D}_5 & \mathbb{Z}_2^2 & 192 & \text{A}_1 \text{A}_2^4 \text{F}_2 & \mathbb{Z}_3 & 239 & \text{C}_9 \text{F}_2 & \text{} \\
\hline
	\end{tabular}
	
	\begin{tabular}
		{
			|>{\columncolor{Gray}$}c<{$}>{\columncolor{white}$}l<{$}>{\columncolor{white}$}c<{$}
			|>{\columncolor{Gray}$}c<{$}>{\columncolor{white}$}l<{$}>{\columncolor{white}$}c<{$}
			|>{\columncolor{Gray}$}c<{$}>{\columncolor{white}$}l<{$}>{\columncolor{white}$}c<{$}
			|>{\columncolor{Gray}$}c<{$}>{\columncolor{white}$}l<{$}>{\columncolor{white}$}c<{$}
			|>{\columncolor{Gray}$}c<{$}>{\columncolor{white}$}l<{$}>{\columncolor{white}$}c<{$}
			|>{\columncolor{Gray}$}c<{$}>{\columncolor{white}$}l<{$}>{\columncolor{white}$}c<{$}
			|>{\columncolor{Gray}$}c<{$}>{\columncolor{white}$}l<{$}>{\columncolor{white}$}c<{$}
			|
		}\hline 
 283 & \text{A}_1 \text{A}_3 \text{A}_4 \text{F}_3 & \mathbb{Z}_2 & 307 & \text{A}_1^2 \text{A}_4 \text{C}_2 \text{F}_3 & \mathbb{Z}_2 & 331 & \text{A}_1 \text{C}_3 \text{D}_4 \text{F}_3 & \mathbb{Z}_2 & 355 & \text{A}_6 \text{F}_2 \text{F}_3 & \text{} & 379 & \text{A}_1 \text{A}_5 \text{C}_1 \text{F}_4 & \text{} & 403 & \text{A}_1 \text{A}_4 \text{F}_2 \text{F}_4 & \text{} \\
 284 & \text{A}_4^2 \text{F}_3 & \text{} & 308 & \text{A}_2 \text{A}_4 \text{C}_2 \text{F}_3 & \text{} & 332 & \text{A}_1 \text{A}_2 \text{D}_5 \text{F}_3 & \mathbb{Z}_2 & 356 & \text{A}_2 \text{D}_4 \text{F}_2 \text{F}_3 & \text{} & 380 & \text{A}_6 \text{C}_1 \text{F}_4 & \text{} & 404 & \text{A}_5 \text{F}_2 \text{F}_4 & \text{} \\
 285 & \text{A}_1^3 \text{A}_5 \text{F}_3 & \mathbb{Z}_2^2 & 309 & \text{A}_1 \text{A}_5 \text{C}_2 \text{F}_3 & \mathbb{Z}_2 & 333 & \text{A}_1^2 \text{C}_1 \text{D}_5 \text{F}_3 & \mathbb{Z}_2 & 357 & \text{A}_1 \text{D}_5 \text{F}_2 \text{F}_3 & \text{} & 381 & \text{A}_1 \text{A}_2^2 \text{C}_2 \text{F}_4 & \text{} & 405 & \text{A}_1 \text{D}_4 \text{F}_2 \text{F}_4 & \text{} \\
\hline
	\end{tabular}\vspace{-.5em}
	\begin{table}[H]
		\caption{Algebras of maximal rank for the heterotic $\mathbb{Z}_2$ triple.}
		\label{tab:algebrasZ2}
	\end{table}
}
\vspace{-.7em}
{\centering\scriptsize
	\setlength{\tabcolsep}{1pt}%
	\setlength\arrayrulewidth{1pt}%
	\begin{tabular}
		{
			|>{\columncolor{Gray}$}c<{$}>{\columncolor{white}$}l<{$}>{\columncolor{white}$}c<{$}
			|>{\columncolor{Gray}$}c<{$}>{\columncolor{white}$}l<{$}>{\columncolor{white}$}c<{$}
			|>{\columncolor{Gray}$}c<{$}>{\columncolor{white}$}l<{$}>{\columncolor{white}$}c<{$}
			|>{\columncolor{Gray}$}c<{$}>{\columncolor{white}$}l<{$}>{\columncolor{white}$}c<{$}
			|>{\columncolor{Gray}$}c<{$}>{\columncolor{white}$}l<{$}>{\columncolor{white}$}c<{$}
			|>{\columncolor{Gray}$}c<{$}>{\columncolor{white}$}l<{$}>{\columncolor{white}$}c<{$}
			|>{\columncolor{Gray}$}c<{$}>{\columncolor{white}$}l<{$}>{\columncolor{white}$}c<{$}
			|>{\columncolor{Gray}$}c<{$}>{\columncolor{white}$}l<{$}>{\columncolor{white}$}c<{$}
			|>{\columncolor{Gray}$}c<{$}>{\columncolor{white}$}l<{$}>{\columncolor{white}$}c<{$}
			|
		}\hline
 1 & \text{A}_1^3 \text{A}_2^2 & \mathbb{Z}_3 & 7 & \text{A}_2 \text{A}_5 & \mathbb{Z}_3 & 13 & \text{A}_1 \text{D}_6 & \text{} & 19 & \text{A}_3^2 \text{G}_1 & \text{} & 25 & \text{A}_1 \text{D}_5 \text{G}_1 & \text{} & 31 & \text{A}_2 \text{A}_3 \text{G}_1^2 & \text{} & 37 & \text{D}_5 \text{G}_1^2 & \mathbb{Z}_2 & 43 & \text{A}_1 \text{D}_4 \text{G}_2 & \text{} & 49 & \text{D}_4 \text{G}_1 \text{G}_2 & \text{} \\
 2 & \text{A}_2^2 \text{A}_3 & \mathbb{Z}_3 & 8 & \text{A}_1 \text{A}_6 & \text{} & 14 & \text{D}_7 & \text{} & 20 & \text{A}_1^2 \text{A}_4 \text{G}_1 & \text{} & 26 & \text{D}_6 \text{G}_1 & \text{} & 32 & \text{A}_2 \text{A}_3 \text{G}_1^2 & \mathbb{Z}_2 & 38 & \text{A}_1^3 \text{A}_2 \text{G}_2 & \text{} & 44 & \text{D}_5 \text{G}_2 & \text{} & 50 & \text{A}_1^3 \text{G}_2^2 & \text{} \\
\hline
	\end{tabular}\vspace{-.5em}
	\begin{table}[H]
		\caption{Algebras of maximal rank for the heterotic $\mathbb{Z}_3$ triple.}
		\label{tab:algebrasZ3}
	\end{table}
}
\vspace{-.7em}
{\centering\scriptsize
	\setlength{\tabcolsep}{1pt}%
	\setlength\arrayrulewidth{1pt}%
	\begin{tabular}
		{
			|>{\columncolor{Gray}$}c<{$}>{\columncolor{white}$}l<{$}>{\columncolor{white}$}c<{$}
			|>{\columncolor{Gray}$}c<{$}>{\columncolor{white}$}l<{$}>{\columncolor{white}$}c<{$}
			|>{\columncolor{Gray}$}c<{$}>{\columncolor{white}$}l<{$}>{\columncolor{white}$}c<{$}
			|>{\columncolor{Gray}$}c<{$}>{\columncolor{white}$}l<{$}>{\columncolor{white}$}c<{$}
			|>{\columncolor{Gray}$}c<{$}>{\columncolor{white}$}l<{$}>{\columncolor{white}$}c<{$}
			|>{\columncolor{Gray}$}c<{$}>{\columncolor{white}$}l<{$}>{\columncolor{white}$}c<{$}
			|>{\columncolor{Gray}$}c<{$}>{\columncolor{white}$}l<{$}>{\columncolor{white}$}c<{$}
			|>{\columncolor{Gray}$}c<{$}>{\columncolor{white}$}l<{$}>{\columncolor{white}$}c<{$}
			|>{\columncolor{Gray}$}c<{$}>{\columncolor{white}$}l<{$}>{\columncolor{white}$}c<{$}
   |
   }\hline
 1 & \text{A}_1^3 \text{A}_2 & \mathbb{Z}_2 & 3 & \text{A}_1^2 \text{A}_3 & \mathbb{Z}_2 & 5 & \text{A}_2 \text{A}_3 & \mathbb{Z}_2 & 7 & \text{A}_5 & \text{} & 9 & \text{D}_5 & \text{} & 11 & \text{A}_1^2 \text{A}_2 \mathbb{A}_1 & \text{} & 13 & \text{A}_1 \text{A}_3 \mathbb{A}_1 & \text{} & 15 & \text{D}_4 \mathbb{A}_1 & \text{} & 17 & \text{A}_1 \text{A}_2 \mathbb{A}_1^2 & \text{} \\
 2 & \text{A}_1 \text{A}_2^2 & \text{} & 4 & \text{A}_2 \text{A}_3 & \text{} & 6 & \text{A}_1 \text{A}_4 & \text{} & 8 & \text{A}_1 \text{D}_4 & \mathbb{Z}_2 & 10 & \text{D}_5 & \mathbb{Z}_2 & 12 & \text{A}_2^2 \mathbb{A}_1 & \text{} & 14 & \text{A}_4 \mathbb{A}_1 & \text{} & 16 & \text{A}_1^3 \mathbb{A}_1^2 & \text{} & 18 & \text{A}_3 \mathbb{A}_1^2 & \text{}\\ \hline \end{tabular}\vspace{-.5em}
	\begin{table}[H]
		\caption{Algebras of maximal rank for the heterotic $\mathbb{Z}_4$ triple. $\mathbb{A}_1$ denotes to an $\mra_1$ at level 1.}
		\label{tab:algebrasZ4}
	\end{table}
}
\vspace{-.7em}
{
\centering\scriptsize
	\setlength{\tabcolsep}{1pt}%
	\setlength\arrayrulewidth{1pt}%
	\begin{tabular}
		{
			|>{\columncolor{Gray}$}c<{$}>{\columncolor{white}$}l<{$}
			|
		}\hline
		1 & \text{A}_1^3 \\
		2 & \text{A}_1 \text{A}_2 \\
		3 & \text{A}_3 \\
		\hline
	\end{tabular}\vspace{-.5em}
	\begin{table}[H]
		\caption{Algebras of maximal rank for the heterotic $\mathbb{Z}_5$ and $\mathbb{Z}_6$ triples.}
		\label{tab:algebrasZ5and6}
	\end{table}
}

\subsection{Fundamental group generators}
\label{app:tab2}
{\centering\scriptsize
	\setlength{\tabcolsep}{1pt}%
	\setlength\arrayrulewidth{1pt}%
	\begin{tabular}
		{
			|>{\columncolor{Gray}$}c<{$}>{\columncolor{white}$}l<{$}>{\columncolor{white}$}c<{$}>{\columncolor{white}$}r<{$}
				|>{\columncolor{Gray}$}c<{$}>{\columncolor{white}$}l<{$}>{\columncolor{white}$}c<{$}>{\columncolor{white}$}r<{$}
				|>{\columncolor{Gray}$}c<{$}>{\columncolor{white}$}l<{$}>{\columncolor{white}$}c<{$}>{\columncolor{white}$}r<{$}
				|>{\columncolor{Gray}$}c<{$}>{\columncolor{white}$}l<{$}>{\columncolor{white}$}c<{$}>{\columncolor{white}$}r<{$}
			|
		}\hline
 4 & \text{A}_1^2 \text{A}_3^3 \text{A}_4^2 & 
\begin{array}{@{}c@{}}
 \mathbb{Z}_2 \\
\end{array}
 & 
\begin{array}{@{}c@{}c@{}c@{}c@{}c@{}c@{}c@{}}
 1 & 1 & 2 & 2 & 2 & 0 & 0 \\
\end{array}
 & 172 & \text{A}_1^3 \text{A}_3 \text{A}_4 \text{A}_9 & 
\begin{array}{@{}c@{}}
 \mathbb{Z}_2 \\
\end{array}
 & 
\begin{array}{@{}c@{}c@{}c@{}c@{}c@{}c@{}}
 0 & 0 & 1 & 2 & 0 & 5 \\
\end{array}
 & 342 & \text{A}_1^2 \text{A}_2 \text{A}_4 \text{A}_7 \text{D}_4 & 
\begin{array}{@{}c@{}}
 \mathbb{Z}_2 \\
\end{array}
 & 
\begin{array}{@{}c@{}c@{}c@{}c@{}c@{}c@{}}
 1 & 1 & 0 & 0 & 4 & \text{s} \\
\end{array}
 & 455 & \text{A}_1 \text{A}_2^2 \text{A}_5 \text{D}_4 \text{D}_5 & 
\begin{array}{@{}c@{}}
 \mathbb{Z}_2 \\
\end{array}
 & 
\begin{array}{@{}c@{}c@{}c@{}c@{}c@{}c@{}}
 1 & 0 & 0 & 3 & \text{s} & 2 \\
\end{array}
 \\
\hline \end{tabular}

\begin{tabular}
		{
			|>{\columncolor{Gray}$}c<{$}>{\columncolor{white}$}l<{$}>{\columncolor{white}$}c<{$}>{\columncolor{white}$}r<{$}
				|>{\columncolor{Gray}$}c<{$}>{\columncolor{white}$}l<{$}>{\columncolor{white}$}c<{$}>{\columncolor{white}$}r<{$}
				|>{\columncolor{Gray}$}c<{$}>{\columncolor{white}$}l<{$}>{\columncolor{white}$}c<{$}>{\columncolor{white}$}r<{$}
				|>{\columncolor{Gray}$}c<{$}>{\columncolor{white}$}l<{$}>{\columncolor{white}$}c<{$}>{\columncolor{white}$}r<{$}
			|
		}\hline
 527 & \text{A}_1^2 \text{A}_{11} \text{D}_6 & 
\begin{array}{@{}c@{}}
 \mathbb{Z}_2 \\
\end{array}
 & 
\begin{array}{@{}c@{}c@{}c@{}c@{}}
 0 & 0 & 6 & \text{v} \\
\end{array}
 & 626 & \text{A}_1 \text{A}_5 \text{D}_6 \text{D}_7 & 
\begin{array}{@{}c@{}}
 \mathbb{Z}_2 \\
\end{array}
 & 
\begin{array}{@{}c@{}c@{}c@{}c@{}}
 1 & 3 & \text{v} & 2 \\
\end{array}
 & 711 & \text{A}_3 \text{A}_7 \text{D}_9 & 
\begin{array}{@{}c@{}}
 \mathbb{Z}_2 \\
\end{array}
 & 
\begin{array}{@{}c@{}c@{}c@{}}
 2 & 4 & 2 \\
\end{array}
 & 798 & \text{A}_2 \text{D}_5 \text{D}_{12} & 
\begin{array}{@{}c@{}}
 \mathbb{Z}_2 \\
\end{array}
 & 
\begin{array}{@{}c@{}c@{}c@{}}
 0 & 2 & \text{s} \\
\end{array}
 \\
 527' & \text{A}_1^2 \text{A}_{11} \text{D}_6 & 
\begin{array}{@{}c@{}}
 \mathbb{Z}_2 \\
\end{array}
 & 
\begin{array}{@{}c@{}c@{}c@{}c@{}}
 1 & 1 & 6 & 0 \\
\end{array}
 & 628 & \text{A}_1 \text{D}_5 \text{D}_6 \text{D}_7 & 
\begin{array}{@{}c@{}}
 \mathbb{Z}_2 \\
\end{array}
 & 
\begin{array}{@{}c@{}c@{}c@{}c@{}}
 1 & 2 & \text{s} & 2 \\
\end{array}
 & 715 & \text{A}_1 \text{A}_9 \text{D}_9 & 
\begin{array}{@{}c@{}}
 \mathbb{Z}_2 \\
\end{array}
 & 
\begin{array}{@{}c@{}c@{}c@{}}
 1 & 5 & 2 \\
\end{array}
 & 799 & \text{A}_1 \text{D}_6 \text{D}_{12} & 
\begin{array}{@{}c@{}}
 \mathbb{Z}_2 \\
\end{array}
 & 
\begin{array}{@{}c@{}c@{}c@{}}
 0 & \text{v} & \text{s} \\
\end{array}
 \\
\hline \end{tabular}

\begin{tabular}
		{
			|>{\columncolor{Gray}$}c<{$}>{\columncolor{white}$}l<{$}>{\columncolor{white}$}c<{$}>{\columncolor{white}$}r<{$}
				|>{\columncolor{Gray}$}c<{$}>{\columncolor{white}$}l<{$}>{\columncolor{white}$}c<{$}>{\columncolor{white}$}r<{$}
				|>{\columncolor{Gray}$}c<{$}>{\columncolor{white}$}l<{$}>{\columncolor{white}$}c<{$}>{\columncolor{white}$}r<{$}
				|>{\columncolor{Gray}$}c<{$}>{\columncolor{white}$}l<{$}>{\columncolor{white}$}c<{$}>{\columncolor{white}$}r<{$}
			|
		}\hline
 993' & \text{A}_1^2 \text{A}_3 \text{A}_7 \text{E}_7 & 
\begin{array}{@{}c@{}}
 \mathbb{Z}_2 \\
\end{array}
 & 
\begin{array}{@{}c@{}c@{}c@{}c@{}c@{}}
 1 & 1 & 2 & 4 & 0 \\
\end{array}
 & 1086 & \text{A}_1 \text{D}_4 \text{E}_7^2 & 
\begin{array}{@{}c@{}}
 \mathbb{Z}_2 \\
\end{array}
 & 
\begin{array}{@{}c@{}c@{}c@{}c@{}}
 0 & \text{s} & 1 & 1 \\
\end{array}
 & 839 & \text{A}_1 \text{A}_2^2 \text{A}_3 \text{A}_5 \text{E}_6 & 
\begin{array}{@{}c@{}}
 \mathbb{Z}_3 \\
\end{array}
 & 
\begin{array}{@{}c@{}c@{}c@{}c@{}c@{}c@{}}
 0 & 1 & 1 & 0 & 2 & 1 \\
\end{array}
 & 442 & \text{A}_1^3 \text{A}_{11} \text{D}_5 & 
\begin{array}{@{}c@{}}
 \mathbb{Z}_4 \\
\end{array}
 & 
\begin{array}{@{}c@{}c@{}c@{}c@{}c@{}}
 0 & 0 & 1 & 3 & 1 \\
\end{array}
 \\
 996 & \text{A}_1 \text{A}_4 \text{A}_7 \text{E}_7 & 
\begin{array}{@{}c@{}}
 \mathbb{Z}_2 \\
\end{array}
 & 
\begin{array}{@{}c@{}c@{}c@{}c@{}}
 1 & 0 & 4 & 1 \\
\end{array}
 & 1088 & \text{D}_5 \text{E}_7^2 & 
\begin{array}{@{}c@{}}
 \mathbb{Z}_2 \\
\end{array}
 & 
\begin{array}{@{}c@{}c@{}c@{}}
 2 & 1 & 1 \\
\end{array}
 & 842 & \text{A}_2^2 \text{A}_4 \text{A}_5 \text{E}_6 & 
\begin{array}{@{}c@{}}
 \mathbb{Z}_3 \\
\end{array}
 & 
\begin{array}{@{}c@{}c@{}c@{}c@{}c@{}}
 1 & 1 & 0 & 2 & 1 \\
\end{array}
 & 445 & \text{A}_1 \text{A}_2 \text{A}_{11} \text{D}_5 & 
\begin{array}{@{}c@{}}
 \mathbb{Z}_4 \\
\end{array}
 & 
\begin{array}{@{}c@{}c@{}c@{}c@{}}
 1 & 0 & 3 & 1 \\
\end{array}
 \\
\hline \end{tabular}

\begin{tabular}
		{
			|>{\columncolor{Gray}$}c<{$}>{\columncolor{white}$}l<{$}>{\columncolor{white}$}c<{$}>{\columncolor{white}$}r<{$}
				|>{\columncolor{Gray}$}c<{$}>{\columncolor{white}$}l<{$}>{\columncolor{white}$}c<{$}>{\columncolor{white}$}r<{$}
				|>{\columncolor{Gray}$}c<{$}>{\columncolor{white}$}l<{$}>{\columncolor{white}$}c<{$}>{\columncolor{white}$}r<{$}
				|>{\columncolor{Gray}$}c<{$}>{\columncolor{white}$}l<{$}>{\columncolor{white}$}c<{$}>{\columncolor{white}$}r<{$}
			|
		}\hline
 22 & \text{A}_1^3 \text{A}_3^2 \text{A}_5^2 & 
\begin{array}{@{}c@{}}
 \mathbb{Z}_2 \\
 \mathbb{Z}_2 \\
\end{array}
 & 
\begin{array}{@{}c@{}c@{}c@{}c@{}c@{}c@{}c@{}}
 0 & 0 & 1 & 2 & 2 & 0 & 3 \\
 0 & 1 & 0 & 2 & 2 & 3 & 0 \\
\end{array}
 & 493 & \text{A}_1^3 \text{A}_5^2 \text{D}_6 & 
\begin{array}{@{}c@{}}
 \mathbb{Z}_2 \\
 \mathbb{Z}_2 \\
\end{array}
 & 
\begin{array}{@{}c@{}c@{}c@{}c@{}c@{}c@{}}
 0 & 0 & 0 & 3 & 3 & \text{v} \\
 0 & 1 & 1 & 0 & 3 & \text{s} \\
\end{array}
 & 635 & \text{A}_1^3 \text{A}_2 \text{A}_3^2 \text{D}_8 & 
\begin{array}{@{}c@{}}
 \mathbb{Z}_2 \\
 \mathbb{Z}_2 \\
\end{array}
 & 
\begin{array}{@{}c@{}c@{}c@{}c@{}c@{}c@{}c@{}}
 0 & 1 & 1 & 0 & 0 & 2 & \text{s} \\
 1 & 0 & 1 & 0 & 2 & 0 & \text{c} \\
\end{array}
 & 741 & \text{A}_1 \text{A}_3 \text{A}_5 \text{D}_{10} & 
\begin{array}{@{}c@{}}
 \mathbb{Z}_2 \\
 \mathbb{Z}_2 \\
\end{array}
 & 
\begin{array}{@{}c@{}c@{}c@{}c@{}}
 0 & 0 & 3 & \text{s} \\
 1 & 2 & 0 & \text{c} \\
\end{array}
\\
\hline \end{tabular}
}

\hspace*{-0.6em}  
\vbox{
{\centering\scriptsize
	\setlength{\tabcolsep}{1pt}%
	\setlength\arrayrulewidth{1pt}%
\begin{tabular}
		{
			|>{\columncolor{Gray}$}c<{$}>{\columncolor{white}$}l<{$}>{\columncolor{white}$}c<{$}>{\columncolor{white}$}r<{$}
				|>{\columncolor{Gray}$}c<{$}>{\columncolor{white}$}l<{$}>{\columncolor{white}$}c<{$}>{\columncolor{white}$}r<{$}
				|>{\columncolor{Gray}$}c<{$}>{\columncolor{white}$}l<{$}>{\columncolor{white}$}c<{$}>{\columncolor{white}$}r<{$}
			|
		}\hline
 376 & \text{A}_1^2 \text{A}_3^3 \text{D}_4^2 & 
\begin{array}{@{}c@{}}
 \mathbb{Z}_2 \\
 \mathbb{Z}_2 \\
 \mathbb{Z}_2 \\
\end{array}
 & 
\begin{array}{@{}c@{}c@{}c@{}c@{}c@{}c@{}c@{}}
 0 & 0 & 0 & 2 & 2 & \text{s} & \text{s} \\
 0 & 0 & 2 & 0 & 2 & \text{c} & \text{c} \\
 1 & 1 & 0 & 0 & 2 & \text{v} & \text{v} \\
\end{array}
 & 538 & \text{A}_1^2 \text{A}_3 \text{D}_4^2 \text{D}_6 & 
\begin{array}{@{}c@{}}
 \mathbb{Z}_2 \\
 \mathbb{Z}_2 \\
 \mathbb{Z}_2 \\
\end{array}
 & 
\begin{array}{@{}c@{}c@{}c@{}c@{}c@{}c@{}}
 0 & 0 & 2 & \text{s} & \text{s} & \text{v} \\
 0 & 1 & 0 & \text{s} & \text{c} & \text{s} \\
 1 & 0 & 0 & \text{c} & \text{s} & \text{c} \\
\end{array}
 & 560 & \text{A}_1^3 \text{D}_4 \text{D}_6^2 & 
\begin{array}{@{}c@{}}
 \mathbb{Z}_2 \\
 \mathbb{Z}_2 \\
 \mathbb{Z}_2 \\
\end{array}
 & 
\begin{array}{@{}c@{}c@{}c@{}c@{}c@{}c@{}}
 0 & 0 & 0 & \text{s} & \text{s} & \text{s} \\
 0 & 0 & 1 & \text{c} & \text{c} & \text{v} \\
 1 & 1 & 0 & 0 & \text{c} & \text{c} \\
\end{array}
 \\
\hline\end{tabular}}
%\end{table}
}
	\begin{table}[H]
		\caption{Groups of maximal rank with non-trivial fundamental group and their generators for the heterotic string on $T^3$.}
		\label{tab:groupsT3}
	\end{table}

{\centering\scriptsize
	\setlength{\tabcolsep}{1pt}%
	\setlength\arrayrulewidth{1pt}%
	\begin{tabular}
		{
			|>{\columncolor{Gray}$}c<{$}>{\columncolor{white}$}l<{$}>{\columncolor{white}$}c<{$}>{\columncolor{white}$}r<{$}
				|>{\columncolor{Gray}$}c<{$}>{\columncolor{white}$}l<{$}>{\columncolor{white}$}c<{$}>{\columncolor{white}$}r<{$}
				|>{\columncolor{Gray}$}c<{$}>{\columncolor{white}$}l<{$}>{\columncolor{white}$}c<{$}>{\columncolor{white}$}r<{$}
				|>{\columncolor{Gray}$}c<{$}>{\columncolor{white}$}l<{$}>{\columncolor{white}$}c<{$}>{\columncolor{white}$}r<{$}
			|
		}\hline
 4 & \text{A}_2^2 \text{A}_7 & 
\begin{array}{@{}c@{}}
 \mathbb{Z}_2 \\
\end{array}
 & 
\begin{array}{@{}c@{}c@{}c@{}}
 0 & 0 & 4 \\
\end{array}
 & 82 & \text{A}_1 \text{A}_5 \text{C}_1 \text{C}_4 & 
\begin{array}{@{}c@{}}
 \mathbb{Z}_2 \\
\end{array}
 & 
\begin{array}{@{}c@{}c@{}c@{}c@{}}
 1 & 3 & 0 & 0 \\
\end{array}
 & 161 & \text{A}_2 \text{C}_1 \text{C}_2 \text{D}_6 & 
\begin{array}{@{}c@{}}
 \mathbb{Z}_2 \\
\end{array}
 & 
\begin{array}{@{}c@{}c@{}c@{}c@{}}
 0 & 0 & 1 & \text{s} \\
\end{array}
 & 312 & \text{A}_1^2 \text{A}_3 \text{C}_3 \text{F}_3 & 
\begin{array}{@{}c@{}}
 \mathbb{Z}_2 \\
\end{array}
 & 
\begin{array}{@{}c@{}c@{}c@{}c@{}c@{}}
 0 & 1 & 2 & 0 & 1 \\
\end{array}
 \\
 5 & \text{A}_2^2 \text{A}_3^2 \text{C}_1 & 
\begin{array}{@{}c@{}}
 \mathbb{Z}_2 \\
\end{array}
 & 
\begin{array}{@{}c@{}c@{}c@{}c@{}c@{}}
 0 & 0 & 2 & 2 & 0 \\
\end{array}
 & 82' & \text{A}_1 \text{A}_5 \text{C}_1 \text{C}_4 & 
\begin{array}{@{}c@{}}
 \mathbb{Z}_2 \\
\end{array}
 & 
\begin{array}{@{}c@{}c@{}c@{}c@{}}
 1 & 3 & 0 & 1 \\
\end{array}
 & 163 & \text{A}_1 \text{C}_1 \text{C}_3 \text{D}_6 & 
\begin{array}{@{}c@{}}
 \mathbb{Z}_2 \\
\end{array}
 & 
\begin{array}{@{}c@{}c@{}c@{}c@{}}
 1 & 0 & 0 & \text{s} \\
\end{array}
 & 316 & \text{A}_5 \text{C}_3 \text{F}_3 & 
\begin{array}{@{}c@{}}
 \mathbb{Z}_2 \\
\end{array}
 & 
\begin{array}{@{}c@{}c@{}c@{}}
 3 & 0 & 1 \\
\end{array}
 \\
\hline\end{tabular}

\begin{tabular}
		{
			|>{\columncolor{Gray}$}c<{$}>{\columncolor{white}$}l<{$}>{\columncolor{white}$}c<{$}>{\columncolor{white}$}r<{$}
				|>{\columncolor{Gray}$}c<{$}>{\columncolor{white}$}l<{$}>{\columncolor{white}$}c<{$}>{\columncolor{white}$}r<{$}
				|>{\columncolor{Gray}$}c<{$}>{\columncolor{white}$}l<{$}>{\columncolor{white}$}c<{$}>{\columncolor{white}$}r<{$}
				|>{\columncolor{Gray}$}c<{$}>{\columncolor{white}$}l<{$}>{\columncolor{white}$}c<{$}>{\columncolor{white}$}r<{$}
			|
		}\hline
 2 & \text{A}_2 \text{A}_3^3 & 
\begin{array}{@{}c@{}}
 \mathbb{Z}_2 \\
 \mathbb{Z}_2 \\
\end{array}
 & 
\begin{array}{@{}c@{}c@{}c@{}c@{}}
 0 & 0 & 2 & 2 \\
 0 & 2 & 0 & 2 \\
\end{array}
 & 47' & \text{A}_1 \text{A}_3^2 \text{C}_2^2 & 
\begin{array}{@{}c@{}}
 \mathbb{Z}_2 \\
 \mathbb{Z}_2 \\
\end{array}
 & 
\begin{array}{@{}c@{}c@{}c@{}c@{}c@{}}
 0 & 0 & 2 & 1 & 1 \\
 1 & 2 & 0 & 0 & 1 \\
\end{array}
 & 106 & \text{A}_1^2 \text{A}_3 \text{C}_6 & 
\begin{array}{@{}c@{}}
 \mathbb{Z}_2 \\
 \mathbb{Z}_2 \\
\end{array}
 & 
\begin{array}{@{}c@{}c@{}c@{}c@{}}
 0 & 1 & 0 & 1 \\
 1 & 0 & 2 & 1 \\
\end{array}
 & 158 & \text{A}_1 \text{A}_2 \text{C}_2 \text{D}_6 & 
\begin{array}{@{}c@{}}
 \mathbb{Z}_2 \\
 \mathbb{Z}_2 \\
\end{array}
 & 
\begin{array}{@{}c@{}c@{}c@{}c@{}}
 0 & 0 & 1 & \text{s} \\
 1 & 0 & 0 & \text{c} \\
\end{array}
 \\
\hline \end{tabular}

\vspace{2em}

\begin{tabular}
		{
			|>{\columncolor{Gray}$}c<{$}>{\columncolor{white}$}l<{$}>{\columncolor{white}$}c<{$}>{\columncolor{white}$}r<{$}
				|>{\columncolor{Gray}$}c<{$}>{\columncolor{white}$}l<{$}>{\columncolor{white}$}c<{$}>{\columncolor{white}$}r<{$}
				|>{\columncolor{Gray}$}c<{$}>{\columncolor{white}$}l<{$}>{\columncolor{white}$}c<{$}>{\columncolor{white}$}r<{$}
				|>{\columncolor{Gray}$}c<{$}>{\columncolor{white}$}l<{$}>{\columncolor{white}$}c<{$}>{\columncolor{white}$}r<{$}
			|
		}\hline
 1 & \text{A}_1^2 \text{A}_3^3 & 
\begin{array}{@{}c@{}}
 \mathbb{Z}_2 \\
 \mathbb{Z}_2 \\
 \mathbb{Z}_2 \\
\end{array}
 & 
\begin{array}{@{}c@{}c@{}c@{}c@{}c@{}}
 0 & 0 & 0 & 2 & 2 \\
 0 & 0 & 2 & 0 & 2 \\
 1 & 1 & 0 & 0 & 2 \\
\end{array}
 & 44 & \text{A}_1^4 \text{A}_3 \text{C}_2^2 & 
\begin{array}{@{}c@{}}
 \mathbb{Z}_2 \\
 \mathbb{Z}_2 \\
 \mathbb{Z}_2 \\
\end{array}
 & 
\begin{array}{@{}c@{}c@{}c@{}c@{}c@{}c@{}c@{}}
 0 & 0 & 0 & 0 & 2 & 1 & 1 \\
 0 & 0 & 1 & 1 & 0 & 1 & 1 \\
 1 & 1 & 0 & 1 & 0 & 0 & 1 \\
\end{array}
 & 135 & \text{A}_1^3 \text{C}_2^2 \text{D}_4 & 
\begin{array}{@{}c@{}}
 \mathbb{Z}_2 \\
 \mathbb{Z}_2 \\
 \mathbb{Z}_2 \\
\end{array}
 & 
\begin{array}{@{}c@{}c@{}c@{}c@{}c@{}c@{}}
 0 & 0 & 0 & 1 & 1 & \text{s} \\
 0 & 0 & 1 & 0 & 1 & \text{c} \\
 1 & 1 & 0 & 0 & 0 & \text{s} \\
\end{array}
 & 157 & \text{A}_1^3 \text{C}_2 \text{D}_6 & 
\begin{array}{@{}c@{}}
 \mathbb{Z}_2 \\
 \mathbb{Z}_2 \\
 \mathbb{Z}_2 \\
\end{array}
 & 
\begin{array}{@{}c@{}c@{}c@{}c@{}c@{}}
 0 & 0 & 0 & 1 & \text{s} \\
 0 & 0 & 1 & 0 & \text{c} \\
 1 & 1 & 0 & 0 & \text{v} \\
\end{array}
 \\
\hline \end{tabular}

	\begin{table}[H]
		\caption{Groups of maximal rank with non-trivial fundamental group and their generators for the heterotic $\mathbb{Z}_2$ triple.}
		\label{tab:groupsZ2}
	\end{table}
}

{\centering\scriptsize
	\setlength{\tabcolsep}{1pt}%
	\setlength\arrayrulewidth{1pt}%
	\begin{tabular}
		{
			|>{\columncolor{Gray}$}c<{$}>{\columncolor{white}$}l<{$}>{\columncolor{white}$}c<{$}>{\columncolor{white}$}r<{$}
				|>{\columncolor{Gray}$}c<{$}>{\columncolor{white}$}l<{$}>{\columncolor{white}$}c<{$}>{\columncolor{white}$}r<{$}
				|>{\columncolor{Gray}$}c<{$}>{\columncolor{white}$}l<{$}>{\columncolor{white}$}c<{$}>{\columncolor{white}$}r<{$}
			|>{\columncolor{Gray}$}c<{$}>{\columncolor{white}$}l<{$}>{\columncolor{white}$}c<{$}>{\columncolor{white}$}r<{$}
			|
		}\hline
 17 & \text{A}_1^3 \text{A}_3 \text{G}_1 & 
\begin{array}{@{}c@{}}
 \mathbb{Z}_2 \\
\end{array}
 & 
\begin{array}{@{}c@{}c@{}c@{}c@{}c@{}}
 1 & 1 & 1 & 2 & 1 \\
\end{array}
 & 32 & \text{A}_2 \text{A}_3 \text{G}_1^2 & 
\begin{array}{@{}c@{}}
 \mathbb{Z}_2 \\
\end{array}
 & 
\begin{array}{@{}c@{}c@{}c@{}c@{}}
 0 & 2 & 1 & 1 \\
\end{array}
 & 1 & \text{A}_1^3 \text{A}_2^2 & 
\begin{array}{@{}c@{}}
 \mathbb{Z}_3 \\
\end{array}
 & 
\begin{array}{@{}c@{}c@{}c@{}c@{}c@{}}
 0 & 0 & 0 & 1 & 1 \\
\end{array}
 & 7 & \text{A}_2 \text{A}_5 & 
\begin{array}{@{}c@{}}
 \mathbb{Z}_3 \\
\end{array}
 & 
\begin{array}{@{}c@{}c@{}}
 0 & 2 \\
\end{array}
 \\
 28 & \text{A}_1^3 \text{A}_2 \text{G}_1^2 & 
\begin{array}{@{}c@{}}
 \mathbb{Z}_2 \\
\end{array}
 & 
\begin{array}{@{}c@{}c@{}c@{}c@{}c@{}c@{}}
 0 & 1 & 1 & 0 & 1 & 1 \\
\end{array}
 & 35 & \text{A}_1 \text{D}_4 \text{G}_1^2 & 
\begin{array}{@{}c@{}}
 \mathbb{Z}_2 \\
\end{array}
 & 
\begin{array}{@{}c@{}c@{}c@{}c@{}}
 0 & \text{s} & 1 & 1 \\
\end{array}
 & 2 & \text{A}_2^2 \text{A}_3 & 
\begin{array}{@{}c@{}}
 \mathbb{Z}_3 \\
\end{array}
 & 
\begin{array}{@{}c@{}c@{}c@{}}
 1 & 1 & 0 \\
\end{array}
 & 15 & \text{A}_1 \text{E}_6 & 
\begin{array}{@{}c@{}}
 \mathbb{Z}_3 \\
\end{array}
 & 
\begin{array}{@{}c@{}c@{}}
 0 & 1 \\
\end{array}
 \\
\hline\end{tabular}
	\begin{table}[H]
		\caption{Groups of maximal rank with non-trivial fundamental group and their generators for the heterotic $\mathbb{Z}_3$ triple.}
		\label{tab:groupsZ3}
	\end{table}
}

\renewcommand{\baselinestretch}{2}
{\begin{table}[H]
\centering\scriptsize
	\setlength{\tabcolsep}{1pt}%
	\setlength\arrayrulewidth{1pt}%
	\begin{tabular}
		{
			|>{\columncolor{Gray}$}c<{$}>{\columncolor{white}$}l<{$}>{\columncolor{white}$}c<{$}>{\columncolor{white}$}r<{$}
				|>{\columncolor{Gray}$}c<{$}>{\columncolor{white}$}l<{$}>{\columncolor{white}$}c<{$}>{\columncolor{white}$}r<{$}
				|>{\columncolor{Gray}$}c<{$}>{\columncolor{white}$}l<{$}>{\columncolor{white}$}c<{$}>{\columncolor{white}$}r<{$}
			|>{\columncolor{Gray}$}c<{$}>{\columncolor{white}$}l<{$}>{\columncolor{white}$}c<{$}>{\columncolor{white}$}r<{$}
			|>{\columncolor{Gray}$}c<{$}>{\columncolor{white}$}l<{$}>{\columncolor{white}$}c<{$}>{\columncolor{white}$}r<{$}
            |
		}\hline
 1 & \text{A}_1^3 \text{A}_2 & \begin{array}{@{}c@{}} \mathbb{Z}_2 \\\end{array}& 
 \begin{array}{@{}c@{}c@{}c@{}c@{}} 0 & 1 & 1 & 0 \\\end{array}& 
 5 & \text{A}_2 \text{A}_3 & \begin{array}{@{}c@{}} \mathbb{Z}_2 \\\end{array}&
 \begin{array}{@{}c@{}c@{}} 0 & 2 \\\end{array} & 
  10 & \text{D}_5 & \begin{array}{@{}c@{}} \mathbb{Z}_2 \\\end{array} & 
\begin{array}{@{}c@{}} 2 \\\end{array}&
%\\ \hline  
 3 & \text{A}_1^2 \text{A}_3 & \begin{array}{@{}c@{}} \mathbb{Z}_2 \\\end{array} & 
\begin{array}{@{}c@{}c@{}c@{}} 0 & 0 & 2 \\\end{array} & 
  8 & \text{A}_1 \text{D}_4 & \begin{array}{@{}c@{}} \mathbb{Z}_2 \\\end{array} & 
\begin{array}{@{}c@{}c@{}} 0 & \text{s} \\\end{array} %& 
% & & &  
\\
\hline 
\end{tabular}
	%\begin{table}[H]
		\caption{Groups of maximal rank with non-trivial fundamental group and their generators for the heterotic $\mathbb{Z}_4$ triple.	
		}
		\label{tab:groupsZ4}
	\end{table}
}

\bibliographystyle{JHEP}
\bibliography{Heterotic7d}

\end{document}